\documentclass[reprint,
superscriptaddress,
nofootinbib,
amsmath,amssymb,
aps,
pra,
]{revtex4-2}
\usepackage{booktabs} 
\usepackage{graphicx}% Include figure files
\usepackage{dcolumn}% Align table columns on decimal point
\usepackage{bm}% bold math
\usepackage{dsfont}
\usepackage{physics}
\usepackage{color}
\usepackage{soul}
\setstcolor{red}
\usepackage{ifthen}
\newboolean{showannotations}   
\setboolean{showannotations}{false} 
\newcommand*{\added}[1]{%
  \ifthenelse{\boolean{showannotations}}{{\color{blue}#1}}{#1}%
}

\newcommand*{\removed}[1]{%
  \ifthenelse{\boolean{showannotations}}{\st{#1}}{}%
}

\newcommand*{\change}[2]{%
  \ifthenelse{\boolean{showannotations}}{{\color{red}\st{#1}}{\color{blue}#2}}{#2}%
}

\begin{document} 

\title{Investigating the Collective Nature of Cavity Modified Chemical Kinetics \\under Vibrational Strong Coupling}
\author{Lachlan P. Lindoy}%
\affiliation{Department of Chemistry, Columbia University, 3000 Broadway, New York, New York, 10027,  U.S.A}
\author{Arkajit Mandal}%
\affiliation{Department of Chemistry, Columbia University, 3000 Broadway, New York, New York, 10027,  U.S.A}
\author{David R. Reichman}
\email{drr2103@columbia.edu}
\affiliation{Department of Chemistry, Columbia University, 3000 Broadway, New York, New York, 10027,  U.S.A}

\begin{abstract}
In this paper we develop quantum dynamical methods capable of treating the dynamics of chemically reacting systems in an optical cavity in the vibrationally strong-coupling (VSC) limit at finite temperatures and in the presence of a dissipative solvent in both the few and many molecule limits.  In the context of two simple models we demonstrate how reactivity in the {\em collective} VSC regime does not exhibit altered rate behavior in equilibrium, but may exhibit resonant cavity modification of reactivity when the system is explicitly out of equilibrium.  Our results suggest experimental protocols that may be used to modify reactivity in the collective regime and point to features not included in the models studied which demand further scrutiny.
\end{abstract}

\maketitle
\section{Introduction} 
Recent experiments suggest that a modification of ground state chemical reactivity can arise via the formation of polaritons (light-matter hybrid quasi-particles) inside infrared (IR) optical cavities in the vibrational strong coupling (VSC) regime where an ensemble of molecular vibrations is coupled to a quantized radiation field~\cite{ThomasS2019, ThomasACID2016, Wonmi2023S, NagarajanJACS2021,MandalCR2023, LatherACID2019, LatherCS2022, AnoopNp2020}. This effect has the potential to unlock the long-sought ability to inexpensively perform mode-selective chemistry, wherein specific chemical bonds can be formed or cleaved by simply tuning the frequency of the cavity photons. However, recent theoretical efforts have only found limited success in providing a microscopic understanding of this remarkable phenomena~\cite{ LindoyJPCL2022, lindoy2023quantum, schafer2022shining, LiNC2021, LiJPCL2021, MandalJCP2022, MatthewJPCC2023,campos2023swinging, mondal2022dissociation, sun2023modification, fischer2023cavity, fischer2023beyond, YingJCP2023,PhilbinJPCC2022, FiechterJCPL2023}, and controversy exist with respect to experimental reproducibility~\cite{ImperatoreJCP2021, WiesehanJCP2021}, impeding further progress. 

Cavity-modified chemical reactivity in the VSC regime has two distinguishing characteristics. The first is that the chemical rate is only strongly modified (enhanced or suppressed) when the photon frequency is close to some characteristic molecular vibration frequency. This is marked by a sharp peak (or dip) in the cavity-modified chemical rate constant as a function of the cavity photon frequency, with the width of the rate profile matching the width of the IR absorption profile~\cite{ThomasS2019, ThomasACID2016, Wonmi2023S, NagarajanJACS2021,MandalCR2023, LatherACID2019, LatherCS2022, AnoopNp2020}. The second is that such cavity modifications operate in the collective regime, where a macroscopic number of molecular vibrations are collectively coupled to the cavity radiation. Consequently, the light-matter coupling between each individual molecules and the cavity radiation is effectively minuscule. Thus, a crucial question arises: {\it Can collective light-matter coupling, which couples cavity radiation and molecules in a delocalized fashion, lead to a modification of chemical reactivity which operates locally?}

Our previous theoretical work~\cite{lindoy2023quantum}, corroborated by recent work~\cite{YingJCP2023, FiechterJCPL2023} demonstrates that a sharp resonant modification of cavity-modified chemical rate stems from the quantum dynamical interplay between the cavity photon mode and molecular vibrations. However, these works operate at the {\it single} molecule level, where an individual molecule is assumed to strongly couple to the cavity radiation mode. This is achieved by artificially scaling the single molecular coupling by $\sqrt{N}$ (where $N$ is the number of molecules coupled to cavity radiation in an experiment) such that the Rabi splitting observed in the single molecule-cavity setup is similar to that of the experiments. While such large single molecular coupling may be achieved in plasmonic cavities~\cite{MondalJPCL2022}, this situation is not representative of the present experiments showing cavity-modified ground state chemistry~\cite{ThomasS2019, ThomasACID2016, Wonmi2023S, NagarajanJACS2021,MandalCR2023, LatherACID2019, LatherCS2022, AnoopNp2020} that operate in the collective regime.

In this work, with fully quantum dynamical simulations, we investigate the role of collective cavity coupling in the cavity-modified chemical kinetics of two different model systems  under various initial conditions. %The formalism we employ is exact in the limit of $N\rightarrow \infty$~\cite{Mori_2013,CarolloPRL2021}. 
Specifically, we explore two initial conditions that correspond to two significantly different physical scenarios. One is when the initial condition is {\it uncorrelated}, where the $N$ molecules are thermalized in the absence of the cavity and at time $t=0$ molecule-cavity interactions are introduced. The other is when the initial condition is {\it correlated}, where the $N$ molecules and the cavity mode are thermalized in the presence of a dissipative environment, { which at zero temperature corresponds to the polaritonic ground state. }

\begin{figure}[!ht]
    \centering
    \includegraphics[width=\columnwidth]{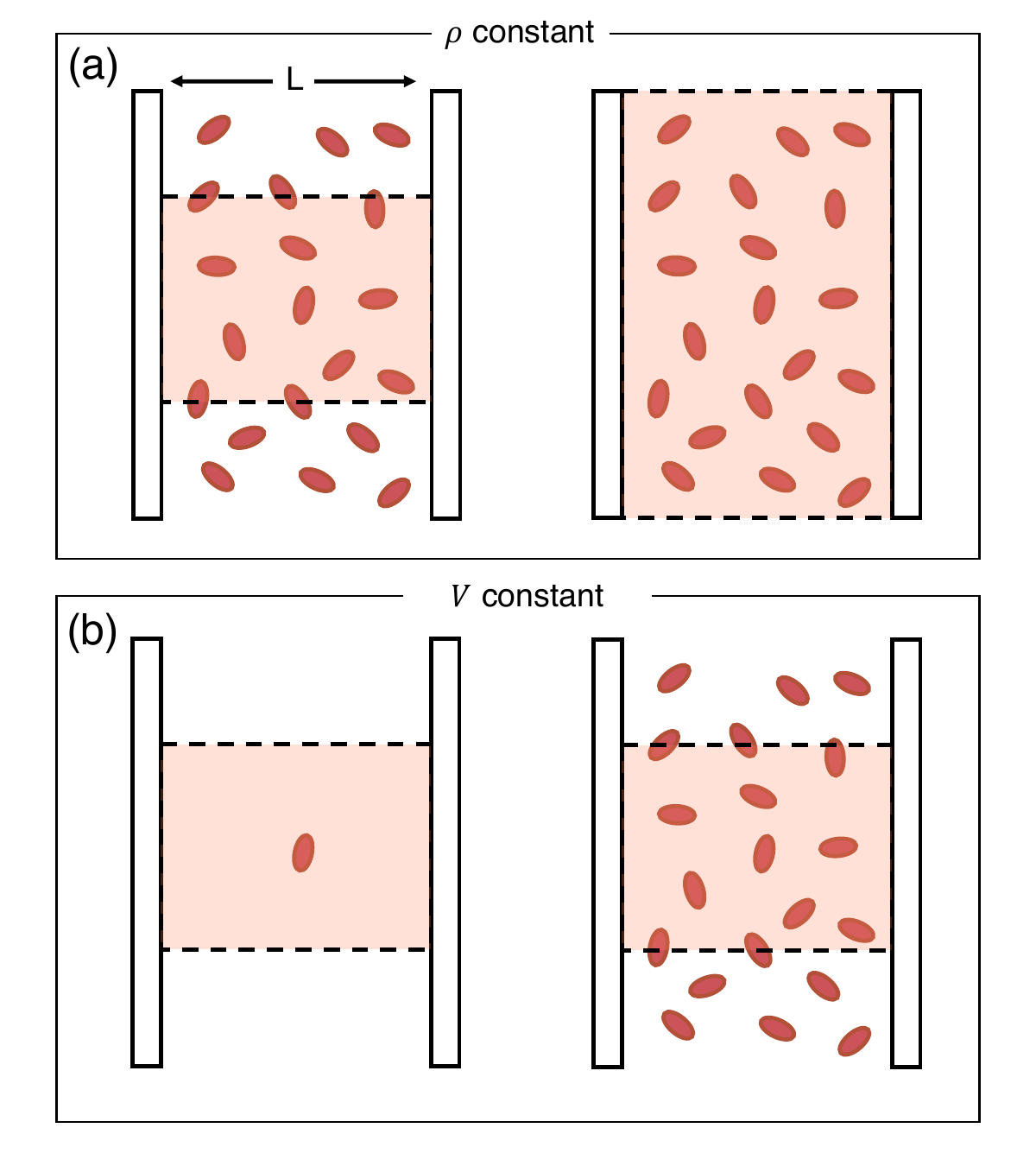}
    \caption{A graphical representation of two possible ways for increasing the number of molecules included in the cavity QED system.  (a) We consider a system at a fixed density $\rho$ and simply increase the volume of the simulation cell (thereby increasing the number of particles included but also increasing the volume).  (b) Treating a fixed simulation volume but varying the number of particles present in it (i.e. changing the density of particles in the cavity by altering $N$). This is the approach taken in references \cite{DerekArxiv2022, PhilbinJPCL2023}}
    \label{fig:n_dependence}
\end{figure}

Our results indicate that under uncorrelated initial conditions, the non-equilibrium decay of state population can be significantly modified when {\it collectively} and {\it resonantly} coupling molecules to the cavity mode, which corroborates previous works that operate in the classical regime~\cite{DerekArxiv2022} or those that operate at zero temperature but treat light-matter interactions quantum mechanically~\cite{fischer2023cavity, JuanPNAS2023}\footnote{While this work was ongoing we became aware of Ref.~\cite{JuanPNAS2023}. The methodology presented here generalizes that of Ref.~\cite{JuanPNAS2023} to finite temperature, many-state systems relevant for modelling of chemical reactions.}.  We find that under such circumstances, nonequilibrium relaxation dynamics is modified, pointing to a possible route to achieving mode-selective chemistry inside optical cavities.  On the other hand, under equilibrium (correlated) initial conditions, we find that chemical reaction rates associated with a barrier crossing process, can be substantially modified only in the few-molecule limit, where the other molecules provide an effective source of cavity dissipation, leading to an enhancement of chemical reaction rate in the energy diffusion-limited regime.  This modification is observed when the cavity mode is resonantly coupled to molecular vibrations. However, we find that this effect is negligible in the mean-field $N \rightarrow \infty$ limit. At the same time, we find that chemical reactions that occur via direct nuclear tunnelling can be modified resonantly where it is found that the coherent oscillations are resonantly damped when coupled to the cavity. Overall, these results point to the possibility of modifying chemical dynamics by coupling molecular vibrations under non-equilibrium conditions while narrowing down the number of factors that might govern the behavior of cavity-modified ground state chemistry. 

This paper is organized as follows: In Section~\ref{sec2} we describe the model systems considered in this study and the quantum dynamics methods that we used in this work. In Section~\ref{sec3} we present and discuss our numerical results. Finally, in Section~\ref{sec4} we document the conclusions of this work and provide avenues for future investigations. 

\section{Theory}\label{sec2} 

\subsection{Model System} 
In this work we consider a minimal model of a set of $N$ molecules collectively coupling to a single radiation mode supported by a cavity.  Fig. \ref{fig:n_dependence}a (see left panel) shows an example of a setup where $N$ molecules in a simulation cell (red shaded area) are placed inside a cavity with width $L$.  There are two ways to analyze the quantum dynamics in the thermodynamic limit: $A \rightarrow \infty$ and $N\rightarrow\infty$, and these two approaches are illustrated in Fig. \ref{fig:n_dependence}a-b.

The Hamiltonian describing the set of molecular vibrations coupled to cavity radiation is written as
\begin{equation}
\hat{H} = \hat{H}_M + \hat{H}_{RM} + \hat{H}_C,
\end{equation}
where $\hat{H}_M$ is the matter Hamiltonian, $\hat{H}_{RM}$ is the radiation-matter coupling Hamiltonian and $\hat{H}_C$ is the Hamiltonian for the cavity radiation mode.  In the single-mode approximation of a perfect lossless optical cavity, the cavity Hamiltonian becomes 
\begin{equation}
    \hat{H}_C = \omega_c \hat{b}_c^\dagger \hat{b}_c.
\end{equation}
The presence of cavity loss can be modelled with a simple Caldeira-Leggett Hamiltonian~\cite{CALDEIRA1983374}, where $\hat{H}_C$ is written as 
\begin{equation}\label{eqn:loss-bath}
    \hat{H}_C = \omega_c \hat{b}_c^\dagger \hat{b}_c + \sum_k \frac{\hat{\Pi}_{k}^{2}}{2} + \frac{1}{2}\tilde{\omega}_{k}^{2} \Big(\hat{\mathcal{Q}}_k + \frac{\mathcal{C}_{k} \hat{q}_c}{\tilde{\omega}_k^{2}}\Big)^2, 
\end{equation}
with $\hat{q}_c = (\hat{b}_c^\dagger + \hat{b}_c)/(\sqrt{2\omega_c})$, and with $\mathcal{C}_{k}$ and $\tilde{\omega}_{k}$  sampled from a spectral density function $J_{loss}(\omega) = \frac{\pi}{2}\sum_{k} \frac{\mathcal{C}_{k}^2}{\tilde{\omega}_{k}^2} \delta(\tilde{\omega}_{k} - {\omega})$ that is taken to be of the Debye form, approximating the result for a simple 1D model cavity~\cite{keeling_notes}.

%\subsubsection{Light Matter Interaction Hamiltonian}
{ In the long-wavelength and single cavity mode limit } the light-matter interaction Hamiltonian can be written as
\begin{equation}
\hat{H}_{RM} = \eta_c \omega_c (\hat{b}_c^\dagger + \hat{b}_c) \boldsymbol{e} \cdot \sum_i \hat{\boldsymbol{\mu}}_i, 
\end{equation}
where $\boldsymbol{e}$  and $\omega_c$ are the polarization direction and frequency of the radiation mode, respectively, 
\begin{equation}\label{light-matter-coup}
    \eta_c = \sqrt{\frac{\hbar}{2\epsilon_0 V\omega_c}},
\end{equation}
with $V$ the quantization volume, and we take the dipole moment operator of molecule $i$ as 
\begin{equation}
    \hat{\boldsymbol{\mu}}_i = \boldsymbol{n}_i \hat{\mu}_i, 
\end{equation}
with $\boldsymbol{n}_i$ a normalized vector that specifies the orientation of molecule $i$.  Here $\eta_c$ quantifies the light-matter interaction strength and may depend on $N$ depending on how we choose to increase the number of molecules present in the system. { Note that the light-matter interactions described in Eq.~\ref{light-matter-coup} ignores the spatial dependence of the radiation field (long-wavelength approximation) which may break down when considering a large number of molecules filling the entirety of the optical cavity. 
}

As mentioned above, Fig. \ref{fig:n_dependence} illustrates two possible strategies to analyze the effect of increasing the number of molecules present in the simulations.  The second strategy, illustrated in Fig. \ref{fig:n_dependence}b, is the choice that has been employed in many recent theoretical works (e.g. \cite{DerekArxiv2022,PhilbinJPCL2023}).  Within the context of the first approach, we consider a simulation box with cross-sectional area (perpendicular to the cavity direction) $A$ that contains $N$ molecules, and use the volume of this box as the quantization volume present in $\chi$.  Doing so we find that 
\begin{equation}
\begin{aligned}
    \eta_c &= \sqrt{\frac{\hbar}{2\epsilon_0 A L\omega_c}} = \frac{\lambda}{\sqrt{A}},
\end{aligned}
\end{equation}
where we have used that $2L = \lambda$ and that $\lambda \omega_c = 2\pi c$, and have absorbed all of the constants into $\lambda$.
Introducing an average particle density per unit of cross sectional area (which is a constant in the experimental setup) $\rho = N/A$, we find that $\eta_c$
\begin{equation}
    \eta_c = \frac{\lambda\sqrt{\rho}}{\sqrt{N}}, 
\end{equation}
and so the final light-matter coupling Hamiltonian in this case can be written
\begin{equation}
\hat{H}_{RM} = \frac{g \omega_c}{\sqrt{N}} (\hat{b}_c^\dagger + \hat{b}_c) \boldsymbol{e} \cdot \sum_i \hat{\boldsymbol{\mu}}_i, 
\end{equation}
where $g = \lambda\sqrt{\rho}$.  This choice, gives rise to a constant Rabi-splitting as the number of molecules is kept constant.
% We will refer to this case as the Dicke-form, and it is the case relevant to exploring the collective vibrational strong coupling regime.

In the second approach,  we  vary the number of particles and keep the quantisation volume constant while increasing the number of particles present in the volume. As a result, the average particle density per unit of cross-sectional area is constant.  In this case, we write
\begin{equation}
\hat{H}_{RM} = \eta_c \omega_c (\hat{b}_c^\dagger + \hat{b}_c) \boldsymbol{e} \cdot \sum_i \hat{\boldsymbol{\mu}}_i
\end{equation}
where $\eta_c$ is independent of $N$, as has been done in recent work~\cite{DerekArxiv2022,PhilbinJPCL2023, sun2023modification}.  This choice of Hamiltonian gives rise to a Rabi-splitting that increases as $\sqrt{N}$.
%\cite{}.  

 In the absence of direct dipole-dipole interactions, the total Hamiltonians in each of these two cases are provided below. The Hamiltonian for the {\it constant} $\rho$ scenario is written as
\begin{align}\label{dse-cross}
\hat{H} &= \hat{H}_M +  \omega_c \hat{b}_c^\dagger \hat{b}_c  \nonumber\\
&+ \frac{g \omega_c}{\sqrt{N}} (\hat{b}_c^\dagger + \hat{b}_c)\sum_i \boldsymbol{e} \cdot \hat{\boldsymbol{\mu}}_i + \frac{g^2 \omega_c}{N}\sum_{ij}\hat{ \boldsymbol{\mu}}_i\cdot\hat{\boldsymbol{\mu}}_j
\end{align}%
with $\hat{H}_M$ describing the bare molecular Hamiltonian. On the other hand, the Hamiltonian for the {\it constant} $V$ approach is written as
\begin{align}\label{dse-cross-V}
\hat{H} &= \hat{H}_M + \omega_c \hat{b}_c^\dagger \hat{b}_c \nonumber \\
&+ \eta_c \omega_c (\hat{b}_c^\dagger + \hat{b}_c)\sum_i \boldsymbol{e} \cdot \hat{\boldsymbol{\mu}}_i + \eta_c^2 \omega_c\sum_{ij} \hat{ \boldsymbol{\mu}}_i\cdot\hat{\boldsymbol{\mu}}_j 
\end{align}
 We note that the inter-molecular dipole self-energy term is known to cancel with the direct Coulomb interaction term when considering {\it all} radiation modes in the light matter Hamiltonian beyond the long-wavelength approximation~\cite{Thirunamachandran1998}.  However, recent work ~\cite{Bernardis2018PRA, MandalNL2023} suggest that when performing a radiation mode truncation, as is done here (considering a single radiation mode), this term should be explicitly included within the dipole gauge Hamiltonian~\cite{TaoACIE2021, TaoPNAS2020}. We also find that this term does not contribute to resonantly modifying the chemical reaction rate.

In this work, we investigate two model molecular systems. In the first, Model I, we consider a molecular system described by a double-well potential energy surface with the reaction coordinate coupled to a set of dissipative solvent degrees of freedom. In the second, Model II, we consider a proton transfer reaction where the reaction coordinate is coupled only to a spectator mode. Below, we describe each of these models.

\subsubsection{Model I}
This model is a multi-molecule generalisation of the cavity VSC system considered in our previous work \citep{lindoy2023quantum}. Each molecular system is described by a symmetric double well potential with the reaction coordinate coupled to a dissipative bath for which well-defined chemical rates can be obtained. In this model, each molecule consists of a single reaction coordinate $\hat{R}$ that is bilinearly coupled to an infinite set of harmonic modes, representing solvent degrees of freedom. We consider the few-molecule case of $N=1,2,3,$ and $4$ molecules, as well as the thermodynamic limit of $N \rightarrow \infty$.  Here, for simplicity, we will only consider the case where all $N$ molecules are aligned with the cavity polarization direction, that is $\boldsymbol{n}_i = \boldsymbol{e}$.  ~\footnote{After the calculations in this work were completed, we became aware of Ref.~\citenum{YechemArxiv2023}, which considers $N=2$ molecules using Model I, and the Hamiltonian where V is held constant. }

The molecular Hamiltonian is given by
\begin{equation}
    \hat{H}_M = \hat{H}_\mathrm{R} + \hat{H}_{solv},
\end{equation}
where the reaction coordinate Hamiltonian is given by
\begin{equation}
\hat{H}_\mathrm{R} =  \hat{T}_{R} + V(\hat{R}),
\end{equation}
with $V(\hat{R})$ describing a simple quartic potential
\begin{equation}
    V(\hat{R}) =  \frac{\omega_b^4}{16E_b} \cdot \hat{R}^4 - \frac{1}{2}\omega_b^2\cdot \hat{R}^2,
\end{equation}
where $\omega_b$ and $E_b$ are the barrier frequency and height, respectively. Equivalently, the reaction coordinate Hamiltonian $\hat{H}_\mathrm{R}$ can be  represented using its vibrational states,

\begin{align}\label{Hmol}
\hat{H}_\mathrm{R}&= \sum_i E_{i}|v_{i}\rangle\langle v_{i}|  \nonumber\\
&\equiv \Bar{ E}_0  \Big(|v_{R}\rangle \langle v_{R} |  +  |v_{L}\rangle \langle v_{L}| \Big) + \sum_{i \ge  2} E_{i}|v_{i}\rangle\langle v_{i}|  \nonumber \\
&+ \Delta \Big(|v_{R}\rangle \langle v_{L}| + |v_{L}\rangle \langle v_{R}|\Big), 
\end{align}
where $\{|v_{i}\rangle\}$ are the vibrational eigenstates of the molecular Hamiltonian ($\hat{H}_\mathrm{R}|v_{i}\rangle = E_{i} |v_{i}\rangle$). In the second line, we have introduced localized states $|v_{L}\rangle = \frac{1}{\sqrt{2}}(|v_{0}\rangle + |v_{1}\rangle)$ and $|v_{R}\rangle = \frac{1}{\sqrt{2}}(|v_{0}\rangle - |v_{1}\rangle)$, with an energy $\Bar{ E}_0 = \frac{1}{2}(E_0 + E_1)$ and a coupling $\Delta = \frac{1}{2}(E_{1} - E_{0})$. These states are the localized ground states of the left and the right wells (blue and red wave functions in Fig.\ref{fig:fig2}a), respectively.  

The solvent Hamiltonian is given by
\begin{equation}
\hat{H}_\mathrm{solv} = \sum_j \frac{\hat{P}_{j}^2}{2} + \frac{1}{2}\Omega_{j}^2 \Big(\hat{X}_j + \frac{C_{j} \hat{R}}{\Omega_j^2}\Big)^2, 
\end{equation}
where the frequencies and coupling constants are determined by the spectral density of the bath, here taken to be of the Debye form $J_{U}(\Omega) =\frac{\pi}{2} \sum_{j} \frac{ {C}_j^2}{ {\Omega}_j} \delta(\Omega -{\Omega}_j) = 2{\Lambda}_{s}\Omega{\Gamma}/(\Omega^2+ {\Gamma}^2) = {\eta}_{s}\Omega{\Gamma}^2/(\Omega^2+ {\Gamma}^2) $. { We note that the quantum dynamical approach used here can be extended to simulate a more realistic molecular system by obtaining the spectral densities describing the bi-linear system-bath couplings and beyond (such as square-linear couplings) and the vibrational levels in Eq.~\ref{Hmol} using {\it ab initio} approaches~\cite{GustinPNAS2023, UenoJCTC2020,SakuraiJPCA2011}.} Finally, in this model, the molecular dipole is taken as $\hat{\mu}= \hat{R}$.
%ASKLACHLAN

% Additionally we will consider a cavity loss contribution to the total Hamiltonian of the form
% \begin{align}
% \hat{H}_\mathrm{loss} = \sum_k \frac{\hat{\Pi}_{k}^{2}}{2} + \frac{1}{2}\tilde{\omega}_{k}^{2} \Big(\hat{\mathcal{Q}}_k + \frac{\mathcal{C}_{k} \hat{q}_c}{\tilde{\omega}_k^{2}}\Big)^2,
% \end{align}
% which gives rise to decay of the total photon population in the cavity.  
\subsubsection{Model II}\label{model2}
This model describes a two-dimensional hydrogen-transfer model constructed for thioacetylacetone, developed by Doslic \textit{et al.} \cite{DoslicPCCP1999} which was recently explored in the context of cavity-modified chemical reactivity in a recent work~ \citep{fischer2023cavity}.  Explicitly, the molecular Hamiltonian for this two-dimensional model with a reaction coordinate $\hat{q}$ (describing H-transfer) and a spectator mode $\hat{Q}$ is written as
\begin{equation}
    \hat{H}_M = -\frac{\hbar^2}{2 m_q} \frac{\partial^2}{\partial q^2} - \frac{\hbar^2}{2 m_Q} \frac{\partial^2}{\partial Q^2} + V(\hat{q}, \hat{Q}),
\end{equation}
where $m_q =  1914.028$ a.u. and $m_Q = 8622.241$ a.u. are the corresponding masses. Further, $V(\hat{q}, \hat{Q})$ is the interaction potential written as 
\begin{equation}
V(\hat{q}, \hat{Q}) = V_0(\hat{q}) + \frac{1}{2}m_Q \omega_Q^2 \left(\hat{Q} - \lambda_s(\hat{q})\right)^2,
\end{equation}
where $\omega_Q = 0.0009728$ a.u. is the spectator mode frequency and $\lambda_s(\hat{q}) = a_q \hat{q}^2 + b_q \hat{q}^3$ with $a_q  = 0.794$ a.u. and $b_q = -0.2688$ a.u. Here $V_0(\hat{q})$ is the reaction path potential and is written as,
\begin{align}
V_0(\hat{q}) = &\frac{V_{\mathrm{OH}}(\hat{q}) + V_{\mathrm{SH}}(\hat{q})}{2} \nonumber \\
&-  \frac{\sqrt{(V_{\mathrm{OH}}(\hat{q}) - V_{\mathrm{SH}}(\hat{q}))^2 + 4K^2(\hat{q})}}{2},
\end{align}
where $V_j(\hat{q}) = \frac{1}{2} m_j \omega_j^2 (\hat{q} - \hat{q}_j^{0})^2 + \Delta_j$ with $j \in \{ \mathrm{OH}, \mathrm{SH}\}$ and $K(\hat{q}) = k_c \exp(-(\hat{q} - q_0)^2)$ with $k_c = 0.15582$ a.u. and $q_0 =  0.2872$ a.u. The remaining parameters are tabulated in Table~\ref{tab:Table1}. Following Ref.~\citep{fischer2023cavity}, we consider a restricted one-dimensional reaction path Hamiltonian in which we restrict dynamics to the 1D path obtained by minimising $V(\hat{q}, \hat{Q})$ with respect to $\hat{Q}$ at each value of $\hat{q}$.  
 
 \begin{table} [!ht]
\caption{Parameters for Model II (a.u.)}
\begin{tabular}{ccccc} 
~~~~~$j$~~~~ 	& ~~~~~$m_j$~~~~ 	& ~~~~~$\omega_j$~~~~  & ~~~~~$q_j^0$~~~~ & ~~~~~$\Delta_j$~~~~\\ \midrule
OH 						& 1728.46 						& 0.01487 & -0.7181 & 0.0				\\
SH						& 1781.32					&  0.01247 &	1.2094 & 	 0.003583			\\
\end{tabular}
\label{tab:Table1}
\end{table}

Following Ref.~\cite{fischer2023cavity}, the dipole moment operator for this one-dimensional model is taken as $\hat{\mu}(\hat{q}) = \mu_0 + \mu_1 (\hat{q} - q_0)$ with $\mu_0 = 1.68$ a.u { and $\mu_1 = -0.129$ a.u}. The reaction coordinate degrees of freedom are discretised using a Colbert-Miller DVR \cite{10.1063/1.462100}, containing 120 grid points, over the range $\hat{q} \in [-1.5, 2.1]a_0$.  The cavity mode is discretised using a harmonic oscillator number basis containing up to $5000$ states depending on the number of molecules, $N$, coupled to the cavity.

% {\bf Model II.} XXXX 
%This corresponds to the unstructured spectral density model considered in our VSC-HEOM paper.  Here we go beyond that work by considering the case of $N = 1,2,3$ molecules and consider the effect of the additional molecules on the reaction kinetics of any given molecule.

% In what follows we will consider the two forms for the light matter coupling strength, namely we will consider the case where $\eta_c$ is independent of $N$ and where $\eta_c \propto N^{-1/2}$.

% what follows we will consider two cases, in section \ref{Sec:CL_collective} a simple model for a symmetric isomerisation reaction in the presence of dissipative interactions with an environment for which well defined chemical rates can be obtained, and in section \ref{Sec:Neq_collective} a more complicated isomerisation reaction involving only a small number of nuclear degrees of freedom for which the closed nature of the system prevents thermalisation to reach a true rate regime.

\subsection{Method} 
\subsubsection{Hierarchical Equations of Motion Approach}\label{HEOM-MF}
 We simulate exact quantum dynamics of the molecular subsystems, each of which are coupled to a dissipative bath, using the Hierarchical Equations of Motion (HEOM) approach.  This well-established open-quantum system dynamics method provides an exact description of the dynamics of a quantum system that is linearly coupled to a set of $N$ harmonic baths \cite{TanimuraJPSP1989,IshizakiJPSJ2005,TanimuraJCP2020}. The details of this method can be found in Ref.~\cite{TanimuraJPSP1989,IshizakiJPSJ2005,TanimuraJCP2020}. In this work we use a customized version of the HEOM approach which is documented in the appendix of our recent paper~\cite{lindoy2023quantum}. We use this approach to simulate the quantum dynamics in Model I when considering cases  with $N=1$ and $2$. For $N = 3$ and $4$, direct solutions of the HEOM is unfeasible, and in this regime we made use of a { Multi-Layer Multiconfiguration Time-Dependent Hartree} (ML-MCTDH) based solver for HEOM \cite{lindoythesis, 10.1063/5.0050720,10.1063/5.0153870}.  Finally, for $N\rightarrow \infty$ we use the mean-field approach described below.

{\bf Mean-Field Quantum Dynamics.}
With increase in $N$, it becomes impractical to perform a direct HEOM calculation of the dynamics of all molecules and their dissipative baths. Instead, we develop a mean-field approach by using the fact that the coupling between each molecule and the cavity mode tends to zero as $1/\sqrt{N}$.  We follow Ref.~\cite{PiperPRL2022} in developing this approach for cavity modified chemical dynamics in the limit $N\rightarrow \infty$, where this approach is expected to be valid~\cite{Mori_2013,CarolloPRL2021,PiperPRL2022}. 
Numerically, we have set $N = 10000$ for obtaining the dynamics. We have verified that the results presented here are converged with respect to $N$.

Within this mean-field treatment we assume that the total density operator of the system comprising $N$ molecules and a single cavity mode can be written in a factorised form for all times
\begin{equation}
\hat{\rho}(t) = \hat{\rho}_c(t) \prod_{i=1}^N \hat{\rho}_i(t).
\end{equation}

Here $\hat{\rho}_c(t)$ and $\hat{\rho}_i(t)$ are the time-dependent density matrices for the cavity radiation mode and the $i$th molecule, respectively. The resultant mean-field equations of motion are  
\begin{align}
\nonumber \frac{\partial}{\partial t} \hat{\rho}_c(t) =& -i\frac{g \omega_c}{\sqrt{N}} \sum_{i=1}^N \mathrm{Tr}\left[\hat{\mu}_i\hat{\rho}_i(t)\right]\left[\hat{b}_c^\dagger + \hat{b}_c, \hat{\rho}_c(t) \right] \\&- i\left[\hat{H}_C, \hat{\rho}_c(t)\right] \\
\nonumber \frac{\partial}{\partial t} \hat{\rho}_i(t) =& -i\frac{g \omega_c}{\sqrt{N}} \mathrm{Tr}\left[(\hat{b}_c^\dagger + \hat{b}_c)\hat{\rho}_c(t)\right]\left[\hat{\mu}_i, \hat{\rho}_i(t) \right] \\ &\hspace{-1.5em}- i\left[\hat{H}_M^i, \hat{\rho}_i(t)\right] -i\frac{2g^2 \omega_c}{N} \sum_{j \neq i} \mathrm{Tr}\left[\hat{\mu}_j\hat{\rho}_j(t)\right]\left[\hat{\mu}_i, \hat{\rho}_i(t)\right].
\end{align}
We observe that for the models of the type considered here, in the mean-field limit, we have a set of $N+1$ coupled equations of motion for each of the individual density operators.  Next, for simplicity we ignore any static disorder in the molecule dipole operators and assume that we start in the initial thermal mean-field state. With these assumptions, we have $\hat{\rho}_i(t) = \hat{\rho}_M(t)$ for all values of $i$.  We  thus reduce this set of $N+1$ coupled equations to a set of two coupled equations:
\begin{align}
\nonumber \frac{\partial}{\partial t} \hat{\rho}_c(t) =& -ig \omega_c\sqrt{N}  \mu_M(t) \left[\hat{b}_c^\dagger + \hat{b}_c, \hat{\rho}_c(t) \right] \\&- i\left[\hat{H}_C, \hat{\rho}_c(t)\right] \\
\nonumber \frac{\partial}{\partial t} \hat{\rho}_M(t) =& -i\frac{g \omega_c}{\sqrt{N}} f_c(t)\left[\hat{\mu}, \hat{\rho}_M(t) \right] \\ &\hspace{-4em}- i\left[\hat{H}_M, \hat{\rho}_M(t)\right] -i\frac{2g^2 \omega_c (N-1)}{N} \mu_M(t)\left[\hat{\mu}, \hat{\rho}_M(t)\right], 
\end{align}
where $\mu_M(t) = \mathrm{Tr}\left[\hat{\mu}_M\hat{\rho}_M(t)\right]$ and $f_c(t) = \mathrm{Tr}\left[(\hat{b}_c^\dagger + \hat{b}_c)\hat{\rho}_c(t)\right]$.  We approximate the dynamics generated by these ODEs by linearising the functions $f_c(t)$ and $\mu_M(t)$ at each time point. Upon doings so, we see that the evolution at each time point is solely the quantum dynamics under a time-dependent Hamiltonian.  Noting that the time-dependent terms only act on the system degrees of freedom in the system-bath model of the molecular Hamiltonian, we apply the HEOM approach to obtain the equations of motion for evolving each of the density operators.  

In Appendix \ref{appendix:mean_field}, we explore the convergence of the mean-field approximation upon increasing the number of molecules, $N$, coupled to the cavity.  These results show that the deviation between the mean-field and description of model I and allow for exploration of the accuracy of the mean-field treatment for large but finite $N$.
%ASKLACHLAN why even have this?
%Alternatively, assuming that we have $N-1$ molecules, $M$, prepared in their thermal state and $1$ molecule, $R$, prepared in the reactant state we need to consider a set of 3 coupled ODEs of the form:
%\begin{align}
%\nonumber \frac{\partial}{\partial t} \hat{\rho}_c(t) =& -ig \omega_c\frac{(N-1)  \mu_M(t) + \mu_R(t)}{\sqrt{N}}  \\&\times \left[\hat{b}_c^\dagger + \hat{b}_c, \hat{\rho}_c(t) \right] - i\left[\hat{H}_C, \hat{\rho}_c(t)\right] \\
%\nonumber \frac{\partial}{\partial t} \hat{\rho}_M(t) =& -i\frac{g \omega_c}{\sqrt{N}} f_c(t)\left[\hat{\mu}, \hat{\rho}_M(t) \right] - i\left[\hat{H}_M, \hat{\rho}_M(t)\right] \\ &\hspace{-2em}-ig^2 \omega_c\frac{ (N-2)\mu_M(t) + \mu_R(t)}{N} \left[\hat{\mu}, \hat{\rho}_M(t)\right], \\
%\nonumber \frac{\partial}{\partial t} \hat{\rho}_R(t) =& -i\frac{g \omega_c}{\sqrt{N}} f_c(t)\left[\hat{\mu}, \hat{\rho}_R(t) \right] - i\left[\hat{H}_R, \hat{\rho}_R(t)\right] \\ &-ig^2 \omega_c\frac{ (N-1)\mu_M(t)}{N} \left[\hat{\mu}, \hat{\rho}_R(t)\right], \\
%\end{align}
%where $\mu_R(t) = \mathrm{Tr}\left[\hat{\mu}_R\hat{\rho}_R(t)\right]$.

{\bf Evaluation of the Forward Reaction Rate.}  In order to evaluate the forward reaction rate, we have assumed that the system is initially in the reactant region with an initial density operator $\hat{\rho}(0) = \hat{\rho}_R$.  The time-dependent reactant and product populations may be written as
\begin{equation}
\begin{aligned}
    P_R(t) &= \mathrm{Tr}\left[(1-\hat{h})\hat{\rho}(t)\right],\\
    P_P(t) &= 1 - P_R(t),
\end{aligned}
\end{equation}
where $\hat{h}$ is the side operator that projects onto the reactant states, and is only a function of the reaction coordinate position operator, $\hat{R}$, in our work.  If first-order kinetics provides a valid description of the reaction process, then in the long-time limit the reactant and product populations will evolve according to the kinetic equations \cite{doi:10.1063/1.5116800, doi:10.1063/5.0098545, QiangJCP2011}
\begin{equation}
\begin{aligned}
\dot{P}_R(t) &= -\kappa P_R(t) + \kappa' P_P(t), \\
    \dot{P}_P(t) &= \kappa P_R(t) - \kappa' P_P(t),
\end{aligned}
\end{equation}
where $\kappa$ and $\kappa'$ are the forward and backward rate constants, respectively (and are related by $\kappa \langle P_R\rangle = \kappa' \langle P_P \rangle$, where $\langle P_R\rangle$ and $\langle P_P \rangle$ are the equilibrium reactant and product populations, which can be obtained from the steady state solution of the HEOM \cite{doi:10.1063/1.4890441}).  Rearranging the expression for the forward rate constant, we have \cite{doi:10.1063/1.2772265, doi:10.1063/1.5116800, QiangJCP2011}
\begin{equation}\label{Eq:RateConstant}
    \kappa = \lim_{t\rightarrow\infty}\frac{\dot{P}_P(t)}{1-P_P(t)/\langle P_P\rangle},
\end{equation}
where the limit $t\rightarrow\infty$ indicates that the kinetic description of the reaction process is only valid after some initial transient process.

In the evaluation of the rate constants, we have considered an initial condition in which the system of $N$ reaction coordinates and a cavity mode are initially uncorrelated from their dissipative baths. Here we have taken an initial density operator of the form 
\begin{equation}
    \hat{\rho}_R = \frac{1}{Z_R}e^{-\beta \hat{H}_S/2} (1-\hat{h})e^{-\beta \hat{H}_S/2} \otimes \frac{e^{-\beta \hat{H}_B}}{\mathrm{Tr}\left[e^{-\beta \hat{H}_B}\right]}, 
\end{equation}
where $Z_R = \mathrm{Tr}\left[e^{-\beta \hat{H}_S/2} (1-\hat{h})e^{-\beta \hat{H}_S/2} \right]$, which allows for the direct application of the HEOM approach.  The short-time transient dynamics depends on the choice of the initial reactant density operator.  For the non-interacting molecule-solvent initial condition, a short-time ($\sim$ 200 fs) transient slippage in the population of the reactant state is observed. However, we have found that the long-time plateau value of Eq.~\ref{Eq:RateConstant} is independent of the choice of initial density operator for the models considered in this work.

\subsubsection{Multi-Layer Multiconfiguration Time-Dependent Hartree Approach}

\begin{figure}[!ht]
    \centering
    \includegraphics[width=\linewidth]{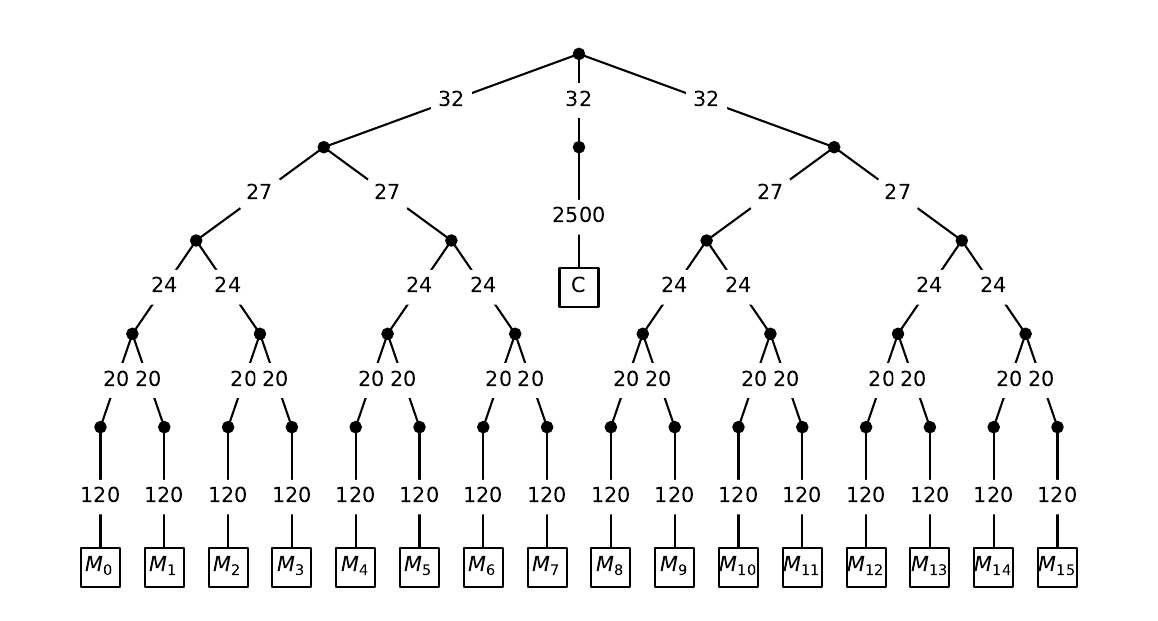}
    \caption{Multilayer tree used for the $N=16$ molecule ML-MCTDH calculations for Model II.  Bond dimension (number of single particle functions) are shown on the edges connecting circular nodes and the primitive Hilbert space dimensions are shown for bonds connecting circular and square nodes.  Here $C$ denotes the cavity degree of freedom and as $M_i$ denotes the $i$th molecules degree of freedom. }
    \label{fig:mctdh_tree}
\end{figure}

For simulating the exact quantum dynamics of Model II, which is a closed quantum system, we use the Multi-Layer Multiconfiguration Time-Dependent Hartree (ML-MCTDH) Approach. ML-MCTDH uses a tensor network based ansatz for the total wavefunction, and employ an efficient projector splitting integrator~\cite{LUBICH2015,KieriSJNA2016,KLOSS2017,BONFANTI2018252,CERUTI2021, lindoy_mctdh_1, lindoy_mctdh_2} that avoids many of the issues that arise in the presence direct product state initial wavefunction conditions \cite{lindoy_mctdh_2}. In all calculations we used a single-site with subspace expansion integration scheme similar to those proposed in reference \cite{MendiveJCP2020} for the standard ML-MCTDH algorithm to allow for growth of the bond-dimension (or number of single-particle functions) throughout the dynamics. In all calculations we consider a ML-MCTDH tree consisting of a balanced binary tree for representing the $N$ molecule system that is connected to the cavity mode at its root. An example of the topology of the ML-MCTDH wavefunction for Model II with $N=16$ is show in Fig. \ref{fig:mctdh_tree}.   Here we find converged results allowing a maximum bond dimension (or number of single particle functions) in the tensor network to be $32$.  The method described below can be generalized to  finite temperature, but for simplicity we work at zero temperature. 

For Model I, we have made use of an ML-MCTDH representation of the HEOM for all simulations with $N>2$.  Here we used the same balanced binary tree structure for the ML-MCTDH tree, with each leaf mode of the molecule representing the full space of auxillary density operators (ADOs) for that molecule and its environment. In this case, we find converged results allowing a maximum bond dimension in the tensor network of $48$.

{\bf Initial State Preparation.} For this model (Model II) molecular system, we explore two types of initial conditions. The first is an uncorrelated initial condition in which the cavity is prepared in the vacuum state in the dipole gauge, as has been employed in recent works on collective VSC~\cite{fischer2023cavity}, 
\begin{equation}\label{eqn:uncorrelated}
\ket{\psi_{uc}}(0) = \ket{0} \bigotimes \ket{\psi_{gs,i}},
\end{equation}
where $\psi_{gs_i}$ is the ground state of the molecule Hamiltonian, $\hat{H}_M$, for molecule $i$, and $\ket{0}$ corresponds to the cavity vacuum state.  Such an initial condition is relevant to the case in which the $N$ molecules are thermalised in the absence of the light-matter coupling and at $t=0$ the light-matter coupling is introduced. 

% Second, we will consider the mean-field ground state obtained by minimizing the energy associated with a mean-field ansatz for the wavefunction.  That is we consider an uncorrelated initial condition, of the form  
% \begin{equation}
% \ket{\psi_{mf}}(0) = \ket{\psi_c} \bigotimes \ket{\psi_{M,i}},
% \end{equation}
% with optimised forms for the cavity and molecule ground states. 

Second, we consider a correlated ground state of the total $N$ molecules and the cavity radiation mode, { e.g. the polaritonic ground state}. This choice for initial condition relates to the case in which interactions between the $N$ molecules and the cavity is introduced at $t\rightarrow-\infty$ and the total system is allowed to evolve under the composite Hamiltonian to reach the ground state at $t=0$ (e.g. the thermalized case).

We obtain the ground state wavefunction using a single-site tree tensor network state optimisation algorithm presented in Ref.~\cite{LarssonJCP2019}, however, with the use of subspace expansion allowing for adaptive control of bond dimension throughout the optimisation.  All simulations presented here were performed using the tree tensor network library ttns$\text{\_}$lib~\cite{Llindoy}.

We report the single-molecule side - total side correlation function
\begin{equation}\label{Eq:side_side}
    \langle\Theta_i(t)\Theta(0)\rangle = \bra{\psi} e^{i\hat{H}t} \Theta_i e^{-i\hat{H}t} \bigotimes_{j=1}^N \Theta_j \ket{\psi},
\end{equation}
where $\Theta_i$ is the side operator projecting onto the reactants for molecule $i$, and $\ket{\psi}$ the initial wave functions described above.  Here we evaluate this quantity by independently evolving two initial wavefunctions, $\ket{\psi_1} = \ket{\psi}$, and $\ket{\psi_2} = \bigotimes_{j=1}^N \Theta_j \ket{\psi}$, and evaluating the matrix elements 
\begin{equation}
    \langle\Theta_i(t)\Theta(0)\rangle = \bra{\psi_1(t)} \Theta_i \ket{\psi_2(t)},
\end{equation}
at each point in time.

\begin{figure*}[!ht]
    \centering
    \includegraphics[width=1.0\linewidth]{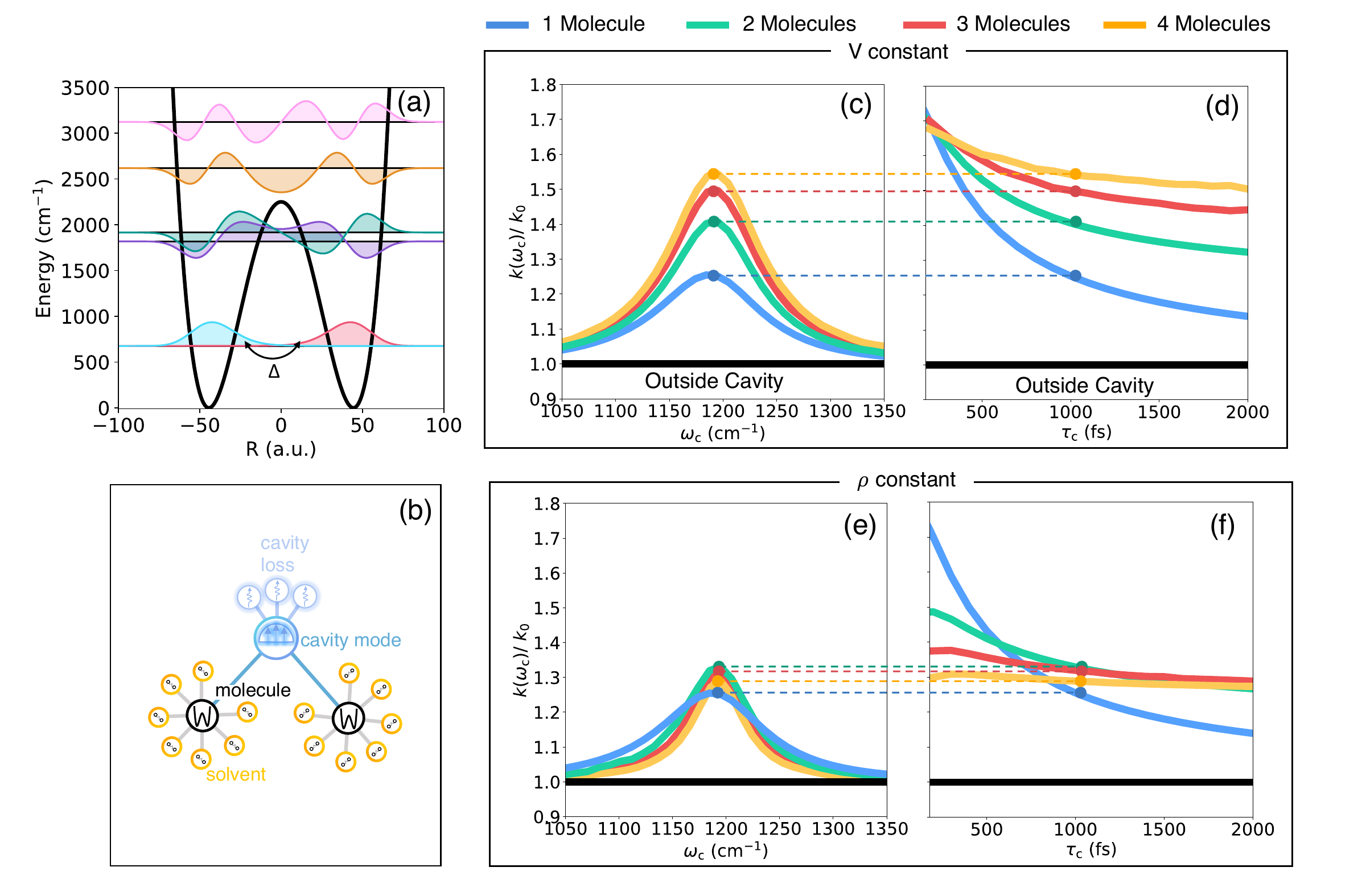}
    \caption{Cavity modified chemical dynamics in the few molecule limit of Model I at T = 300K. (a) Double-well potential describing a model molecular system with the vibrational eigenstates. (b) Schematic illustration of multiple molecular systems coupled to a lossy cavity radiation mode. (c) Cavity photon frequency-dependent normalized chemical reaction rate constant when coupling N = $1, 2, 3$ and 4 molecules to a radiation mode using the constant $V$ approach (see Fig.~\ref{fig:n_dependence}b). (d) Cavity modified chemical rate constant as a function cavity lifetime $\tau_c$ at various $N$. (e)-(f) Same as (c) and (d) but with a constant $\rho$ approach (see Fig.~\ref{fig:n_dependence}a). {For $N=1$ and $2$ molecules the results were obtained using HEOM calculations. For $N=3$ and $4$ molecules the HEOM/ML-MCTDH approach was used.}}
    \label{fig:fig2}
\end{figure*}

\section{Results and Discussion}\label{sec3}
 
{\bf Few Molecule Limit.} We first discuss how a cavity modifies chemical reactivity in the few molecule limit. Via exact quantum dynamics simulations, we find that additional molecules coupled to the cavity radiation mode provide additional dissipation, leading to the enhancement of chemical reactivity for solvent interactions in the energy diffusion-limited regime.  

In Fig. \ref{fig:fig2}, we present the frequency dependent reaction rate, normalised by the out-of-cavity rate, associated with the individual molecules for $N=1,2, 3$ and $4$ molecules in the cavity. As we demonstrated in our recent work~\cite{lindoy2023quantum}, in the single molecule limit, cavity coupling leads to a resonant enhancement of the chemical reaciton rate (blue solid line in Fig.~\ref{fig:fig2}c) due to the additional dissipation originating from the bath that describes cavity loss. This is because the overall chemical reaction rate is limited by the rate of thermal relaxation when the molecule-solvent coupling is of a magnitude such that the reaction falls within the energy diffusion-limited regime. Additional sources of dissipation, such as the dissipation from coupling to a lossy cavity mode, naturally increase the rate of thermalalization, leading to an enhancement of the reaction rate~\cite{lindoy2023quantum}. It is worth noting that the rate increases with the increase in the cavity loss rate (or a decrease in the cavity lifetime) achieved by increasing the coupling between a cavity mode and its bath in Eqn.~\ref{eqn:loss-bath}.

In Fig.~\ref{fig:fig2}c-f, we show that the cavity modification of chemical reaction rate is further increased as the number of molecules in the cavity is increased.  We explore this in Fig.~\ref{fig:fig2}c by keeping the per molecule coupling to the cavity mode a constant, which would correspond to keeping $V$ fixed as illustrated in Fig.~\ref{fig:n_dependence}b. In Fig.~\ref{fig:fig2}c the  cavity enhances the chemical reaction rate by $\sim$25$\%$ at $N =1$ { (Rabi-splitting of $\approx 26.58$ cm$^{-1}$)}. At $N =2, 3$ and 4 ({ with the  Rabi splittings of $37.541$, 45.95  and 53.0 cm$^{-1}$}), the cavity enhancement rises to $\sim$55$\%$. This interesting effect can be explained in terms of enhancement of the overall dissipation. From the perspective of one reactive molecule, additional molecules coupling to cavity can be viewed as additional dissipative bath degrees of freedom that are coupled to the cavity mode thereby increasing cavity loss. However, the extent of this effect quickly saturates and as a result cavity modification of chemical reaction rate marginally changes when going from $N = 3$ to $N=4$. This reveals that even though there is an $N$ dependence of the cavity modified chemical rate, it saturates for a small number of molecules coupled to the cavity.

\begin{figure}[!ht]
    \centering
    \includegraphics[width=0.9\linewidth]{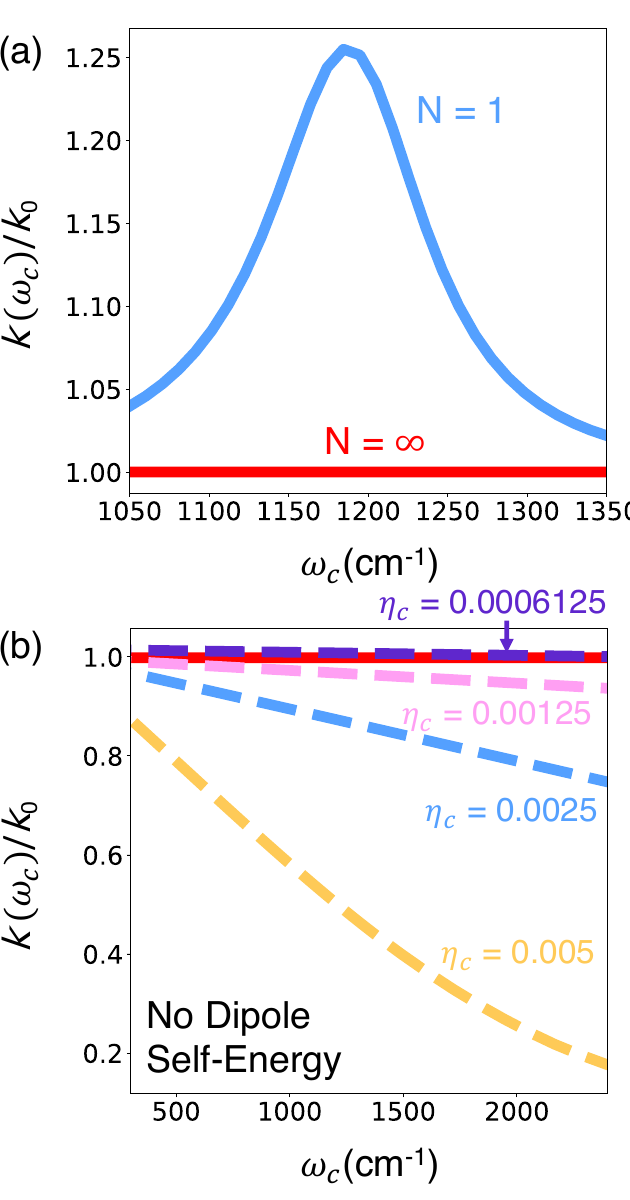}
    \caption{ Cavity modified chemical reactivity in Model I  in the thermodynamic limit at T = $300$ K { obtained using the {\it constant} $\rho$ light-matter coupling Hamiltonian}. (a) Cavity modified (normalized) chemical rate constant as function of photon frequency at $N= 1$ (blue solid line) and at $N = \infty$. (b) Cavity modified (normalized) chemical rate constant as function of photon frequency at $N = \infty$ but in the absence of the dipole self-energy term. }
    \label{fig:fig3}
\end{figure}

Fig.~\ref{fig:fig2}d shows that the extent to which an additional molecule coupling to cavity can enhance chemical reactivity depends on the cavity loss rate. Here we observe that increasing the number of molecules $N$ reduces the sensitivity of the peak rate modification to the cavity loss rate. Intuitively, if the cavity mode coupling strength to its dissipative bath is already very large (i.e. very lossy), the impact of additional sources of dissipation would be comparatively negligible. Similarly, for a nearly perfect lossless cavity, the coupling to an additional molecule can lead to a significant increase in dissipation, and hence a substantial modification of the reaction rate for the reactive molecule. This can be seen in the cavity modified reaction rate as a function of cavity lifetime, as shown in Fig.~\ref{fig:fig2}d. For a small cavity lifetime, $\tau_c <400$ fs, the cavity modified reaction rate is nearly the same for $N = 1, 2, 3$ and $4$. The presence of additional molecule starts to significantly alter chemical reactivity for $\tau_c > 400$ fs.

{ Fig.~\ref{fig:fig2}e-f presents the cavity-modified chemical reaction rate in the presence of a small number of molecules where the total collective coupling is kept constant with a constant Rabi splitting of $26.58$ cm$^{-1}$.} This is achieved by rescaling the per-molecule coupling constant to the cavity mode by a factor of $1/\sqrt{N}$. Thus, the overall cavity modification of the rate occurs via the interplay of two competing effects: the scaling down of individual molecular coupling to the cavity versus the additional dissipation due to the coupling of other molecules to the cavity mode. For the particular solvent couplings chosen here, the cavity modification of the rate is maximized at $N$=2 when using $\tau_c = 1000$ fs (see Fig.~\ref{fig:fig2}e). As before, whether an increase in $N$ will lead to an increase in chemical reactivity also depends on the cavity lifetime. 

Fig.~\ref{fig:fig2}f presents the cavity-modified chemical reaction rate as a function of  the cavity lifetime with $N = 1, 2, 3$  and $4$. We observe that while for larger cavity lifetimes ($\tau_c > 800$ fs)  additional molecules further enhance the chemical reaction rate,  chemical reactivity is less enhanced for smaller cavity lifetimes
 ($\tau_c < 500$ fs). This is because the cavity mode is already strongly connected to a dissipative bath and adding another molecule does not change this dissipation much. As a result, for smaller cavity lifetimes, as the per-molecule coupling is reduced the overall enhancement decreases when increasing the number of molecules. Interestingly, for larger cavity lifetimes ($\tau_c > 800$ fs, the additional dissipation when increasing the number of molecules enhances the chemical reaction rate despite the lower per-molecule light-matter coupling. 

{\bf Thermodynamic  Limit.} We explore  the $N\rightarrow \infty$  using the mean-field HEOM approach explained in Sec.~\ref{HEOM-MF}. As hinted at from the trends observed in Fig.~\ref{fig:fig2}, we expect no cavity modification in the $N\rightarrow\infty$ case.  Numerically, we have set $N = 10000$ ($N \rightarrow \infty$) and have ensured that these results are converged with respect to $N$.

This expectation is confirmed in Fig.~\ref{fig:fig3}a, which compares the cavity-modified chemical kinetics at $N = 1$ (blue solid line) and $N = \infty$. The coupling to the cavity radiation field leads to a resonant modification of chemical kinetics at $N = 1$ and at the same time, this modification vanishes for $N = \infty$ in the mean-field limit (red solid line). Note that here we properly include the dipole self-energy terms (see Eqn.~\ref{dse-cross}) whose inclusion within the light-matter Hamiltonian has been a subject of ongoing debate~\cite{FeistNP2021, Christian2020, TaylorOL2022, RokajJPB2018, GalegoPRX2019, MandalNL2023}.

 Fig.~\ref{fig:fig3}b presents the cavity-modified chemical dynamics in the absence of the dipole self-energy (DSE) terms. In the absence of DSE terms, we do observe a large collective cavity modification of chemical kinetics for $N\rightarrow\infty$. Specifically, we observe a suppression of the chemical kinetics, which is also cavity frequency-dependent. However, we do not observe a resonant effect, where the cavity suppresses chemical reactivity strongly at a certain photon frequency. This observed effect can be explained in terms of the modification of the reaction barrier in the absence of the DSE terms, as explained in Ref.~\cite{GalegoPRX2019}. We observe that this effect increases with the increase in the light-matter coupling strength $\eta_c$.  Note that, in comparison to Ref.~\cite{GalegoPRX2019}, here we explicitly simulate the dynamics of the solvent degrees of freedom.

In Fig.~\ref{fig:fig3} the initial state of the system has been prepared in thermal equilibrium with the molecules placed in the reactant well. Next, we explore the cavity modification of chemical dynamics, where the cavity radiation field is in thermal equilibrium in the absence of the light-matter coupling which is introduced at $t = 0$. Such an initial condition is naturally non-equilibrium and shows different short time dynamics when compared to the dynamics starting with an equilibrated cavity field. 
 
\begin{figure}[!ht]
    \centering
    \includegraphics[width=0.95\linewidth]{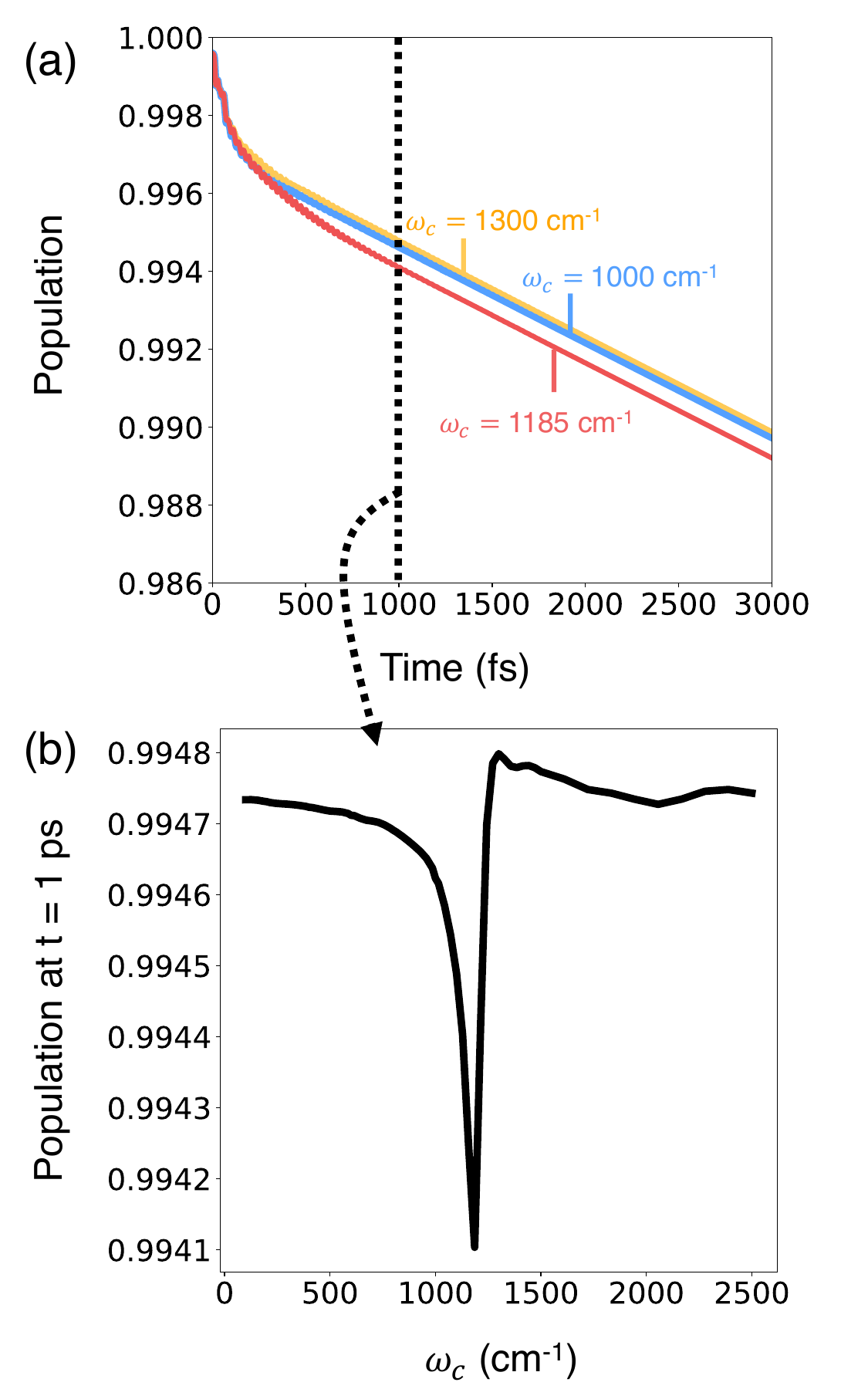}
    \caption{Cavity modified chemical dynamics in Model I under non-equilibrium (uncorrelated) initial conditions at T $300$K. { Results were obtained with the Mean-Field HEOM approach for the {\it constant} $\rho$ light-matter coupling Hamiltonian.} (a) Time-dependent reactant population at various cavity photon frequencies $\omega_c$ with $\omega_c = 1185$ cm$^{-1}$  the resonant frequency. (b) Reactant population at 1 ps as a function of photon frequency, note the size of the effect. }
    \label{fig:fig4}
\end{figure} 

Fig.~\ref{fig:fig4} presents the time-dependent reactant population dynamics when starting from a non-equilibrium initial condition in the thermodynamics limit. The population dynamics presented in Fig.~\ref{fig:fig4}a show a short-time (sub-picosecond) relaxation, which exhibits a photon frequency dependence. When the photon frequency is close to the molecular vibrational transition, the relaxation is faster compared to when the photon frequency is off-resonant. The suppression of the reactant population (or enhancement of chemical reactivity) observed here is due to fact that the cavity radiation is `hotter' than the molecular subsystem at the initial time and this excess heat results in enhanced reactivity at short times. 

At longer times, due to the presence of the solvent bath, the cavity radiation mode as well as the molecular system thermalize, and as a result this leads to chemical dynamics that is same as when starting from a thermalized initial condition. Consequently, the long-time chemical rate constant (compare the slopes of the three curves $t > 1000$ fs), shows no cavity frequency dependence and is identical to the   results in the absence of the cavity. 

Fig.~\ref{fig:fig4}b presents the reactant population at $t = 1$ ps at various photon frequencies. The reactant population shows a sharp cavity resonance feature, signifying a collective and resonant cavity enhancement of chemical reactivity. However, the effect is exceedingly small. The exceedingly small magnitude of this effect reflects the closeness of the non-equilibrium initial condition used here to the equilibrium, thermalized density matrix. While the effects shown in Fig.~\ref{fig:fig4}b are very tiny, they do suggest the tantalizing possibility of modifying chemical reactivity under non-equilibrium scenarios, a topic we return to below and in the conclusions. 

{\bf Cavity-modified Proton Transfer Reaction.}  Below we explore cavity-modified dynamics in a model molecular system describing a proton transfer reaction in thioacetylacetone (see details in Sec.~\ref{model2}). For the following results presented, we define a dimensionless light-matter coupling strength of $\eta = 0.05$, related to the light-matter interaction by
\begin{equation}
    g = \frac{\hbar \eta}{\mu_{10}},
\end{equation}
where $\mu_{10} = \bra{\psi_0}\hat{\mu} \ket{\psi_1} = 0.042$ a.u. \cite{fischer2023cavity}.  { In all calculations for model II, we make use of the {\it constant} $\rho$ form for the light-matter coupling Hamiltonian (Eqn. \ref{dse-cross}). This corresponds to a Rabi-Splitting of $\approx12.59$ cm$^{-1}$ at the resonant photon frequency.}

Here we aim to explore the importance of appropriate initial conditions for the $N$ molecules and the cavity radiation mode when exploring collective VSC.  We consider the dynamics of $N$ molecules interacting with a cavity radiation mode that is resonant with the first vibrational transition of the molecular  Hamiltonian, $\hat{H}_M$, and explore the dependence of the dynamics on $N$ for uncorrelated and correlated (here at $T=0$ K) initial conditions. 

\begin{figure}[!ht]
    \centering
    \includegraphics[width=1.0\linewidth]{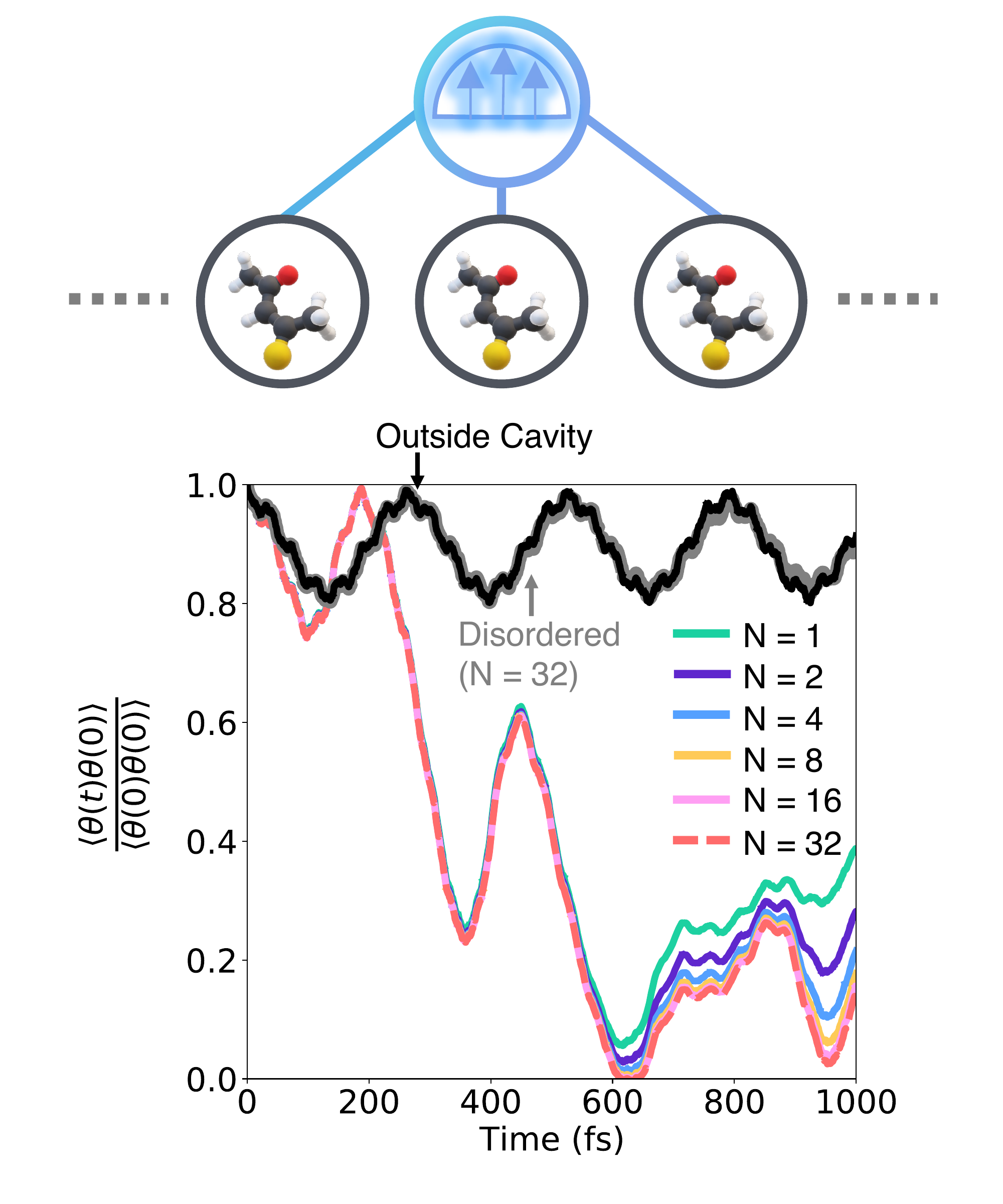}
    \caption{ Cavity-modified proton transfer dynamics in a model thioacetylacetone system { obtained using ML-MCTDH}. The upper panel schematically illustrates how $N$ molecules are coupled to a radiation mode. The bottom panel presents the time-dependent reactant population when starting with a non-equilibrium (uncorrelated) initial condition at various $N$.  Light-matter introductions were included using the {\it constant} $\rho$ light-matter coupling Hamiltonian.}
    \label{fig:fig5}
\end{figure} 

In Fig. \ref{fig:fig5}, we compare the time-dependence of the side-side correlation function defined in Eq. \ref{Eq:side_side} calculated in the absence of a cavity to that with a cavity mode for various values of $N = 1, 2, 4, 8, 16$ and $ 32$.  Here, we have prepared an uncorrelated ground state as defined in Eqn.~\ref{eqn:uncorrelated} that is out of equilibrium (with the light-matter coupling introduced at $t = 0$) and consider a scenario where all the molecules are aligned along the cavity radiation polarization direction. We observe that the presence of a strongly coupled cavity significantly perturbs the coherent crossing dynamics of the molecular system. Fig. \ref{fig:fig5} shows a significant enhancement of the crossing dynamics indicated by the much more rapid decay of the correlation function through the first $100$ fs, as well as a more substantial transfer of population at later times. These results are consistent with those observed in reference \cite{fischer2023cavity} and the molecular dynamics simulations presented in reference \cite{DerekArxiv2022}.  Further, we observe that upon increasing the number of molecules, this dynamics stabilises, with minimal deviations observed between the results obtained with $N=16$ and $32$, even up to times of $1000$ fs. Given the convergence of the results with respect to $N$, we conclude that the observed modification of chemical dynamics occurs in the collective (or in the thermodynamic limit, $N\rightarrow\infty$) regime when using a non-equilibrium initial condition.

{A number of previous works have shown that static and dynamic disorder can play a crucial role in the cavity modified dynamics of molecules and materials~\cite{MichettiPRB2005,SuyabatmazJCP2023, AroeiraNP2024, berghuis2022controlling, Xu2023, QiuJPCL2021, EngelhardtPRL2023, DuPRL2022}.
Here,} in Fig. \ref{fig:fig5} we show that the non-equilibrium effects observed when considering the uncorrelated initial conditions decrease significantly in the presence of angular disorder (see gray solid lines) where molecules are randomly oriented with respect to the cavity radiation polarization direction. We observe that the disordered $N = 32$ scenario shows negligible modification of the mean crossing dynamics when coupled to the cavity radiation, even when starting from a non-equilibrium initial condition. The reason for the disappearance of the cavity modification can be understood by considering the initial (expected) value $\langle {\bf e} \cdot \sum_i \hat{\boldsymbol{\mu}}_i \rangle = 0$ which brings the initial non-equilibrium state close to the correlated ground state, thereby reducing the non-equilibrium effect seen in the crossing dynamics. 
 
Similar to the dynamics in the disordered scenario, the modification of reactivity becomes small when starting from a correlated initial condition corresponding to equilibrium in the reactant well. Interestingly, while the cavity modification of the crossing dynamics is much smaller in this scenario, we do observe a  decay of the oscillation amplitude that exhibits a resonant feature. 

\begin{figure}[!ht]
    \centering
    \includegraphics[width=1.0\linewidth]{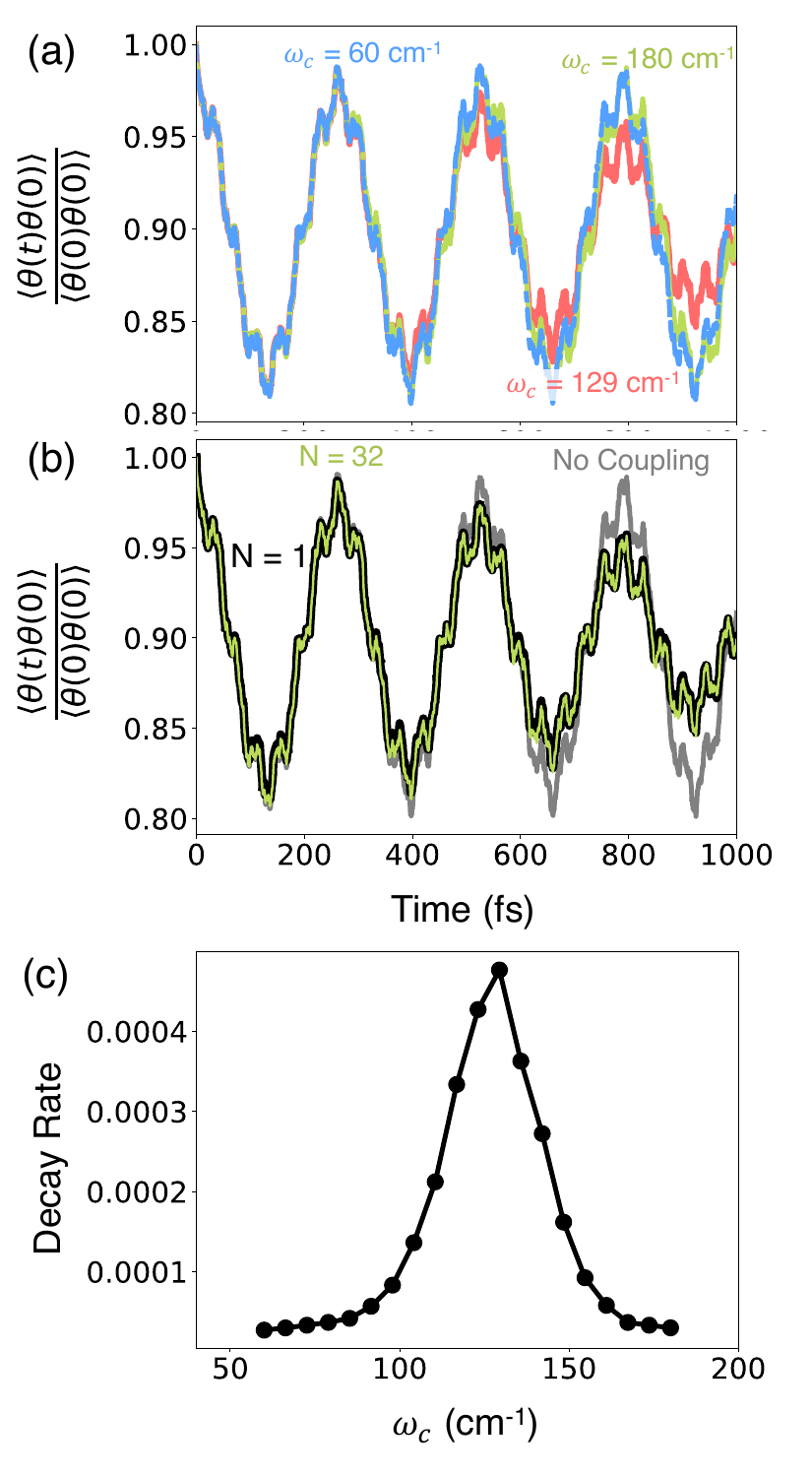}
    \caption{Cavity-modified proton transfer dynamics in a model thioacetylacetone system when starting from a correlated initial condition. (a) Reactant population dynamics at three different photon frequencies with $\omega_c = 129$ cm$^{-1}$ as the resonant photon frequency. (b)  Reactant population dynamics at various $N$ compared to the no coupling scenario. (c) Reactant population decay as a function of the photon frequency obtained for $N=8$. {Results were obtained using ML-MCTDH and using the {\it constant} $\rho$ light-matter coupling Hamiltonian.}}
    \label{fig:fig6}
\end{figure} 

Fig.~\ref{fig:fig6} presents the time-dependent reactant population at three different cavity photon frequencies when starting from a correlated initial condition. We observe that when the cavity photon frequency is in resonance with the vibrational transition, the amplitude of the crossing dynamics is modified. In  Fig.~\ref{fig:fig6}a where the amplitude of the crossing dynamics is largely unmodified for the photon frequency $\omega_c = 60$ cm$^{-1}$ (blue dashed line) or $180$ cm$^{-1}$ (blue solid line). However, when the photon frequency is close to the vibrational transition of the molecular sub-system,
at $\omega_c = 129$ cm$^{-1}$ we observe a decay of the amplitude of the side-side correlator. 

In Fig.~\ref{fig:fig6}b we confirm that this decay persists in the collective limit by comparing the $N = 1$ and $N = 32$ cases.  Fig.~\ref{fig:fig6}b shows that the overall dynamics are identical between the two scenarios. 

To analyze the resonant behavior of the cavity modified dynamics, we compute a decay rate, $\kappa$, obtained by minimising the function
\begin{equation}
f(a,b,\kappa;\omega_c) = \int_0^{t_{\mathrm{max}}} \mathrm{d}t\left|e^{-\kappa t}(\theta_{0}(t) - a) -  (\theta(t; \omega_c) - b) \right|^2,
\end{equation}
where $a, b, \kappa$ are the fitting parameters, $\theta_{0}(t)$ is the normalised side-side correlation function without the cavity and $\theta(t; \omega_c)$ is the side-side correlation function obtained in the presence of a cavity mode.  

Fig.~\ref{fig:fig6}c presents the decay rate $\kappa$ as a function of the photon frequency $\omega_c$. Overall, the decay rate clearly shows a resonant peak with $\omega_c$ close to the vibrational transition of the molecular system. This decay can be understood as a decoherence process due to the presence of the cavity radiation mode, which is acting as a bath  degree of freedom. Note that this decay is transient, and at longer times, the populations will eventually return since this is still a small, closed quantum system. At the same time, we do expect this decay to persist when considering a lossy cavity mode. 

While such cavity-modified dynamics are {\it collective} and {\it resonant}, whether or not such effects can be substantial when considering solvent interactions with the molecular system remains an open question which we briefly return to before concluding. Further, we also note that the damping of the crossing dynamics does not necessarily indicate a modification of chemical reactivity, since the average product population (when averaging over the oscillations) remains the same for the time-range presented here. 

The results in Fig.~\ref{fig:fig5} and Fig.~\ref{fig:fig6} also shed light on reports on the presence or absence of  collective cavity modification of chemical reactivity in recent work~\cite{DerekArxiv2022, SunJPCL2022,MatthewJPCC2023}. Specifically, a number of studies employing classical trajectory-based simulations \cite{DerekArxiv2022, SunJPCL2022}, classical rate theories, \cite{MatthewJPCC2023} and quantum dynamical simulations \cite{fischer2023cavity} have obtained conflicting results concerning vibrational strong coupling in the collective coupling regime.  In Ref.~\citep{DerekArxiv2022,fischer2023cavity}, cavity modification of chemical dynamics is observed that persists to large $N$, in contrast, in Ref.~\citep{SunJPCL2022,MatthewJPCC2023} no modification of the chemical reaction rate is observed in the large $N$ limit.  Our results illustrate that the collective effects observed in previous works arise from the choice of the initial condition in which the $N$ molecules and cavity mode are initially uncorrelated.  We further demonstrate that such collective effects vanish upon the use of an appropriate correlated initial condition  consistent with the linear response definition of rate theory.  While we  have demonstrate this in the zero-temperature case, the underlying argument holds for finite temperatures as well.

\begin{figure}[!ht]
    \centering
    \includegraphics[width=1.0\linewidth]{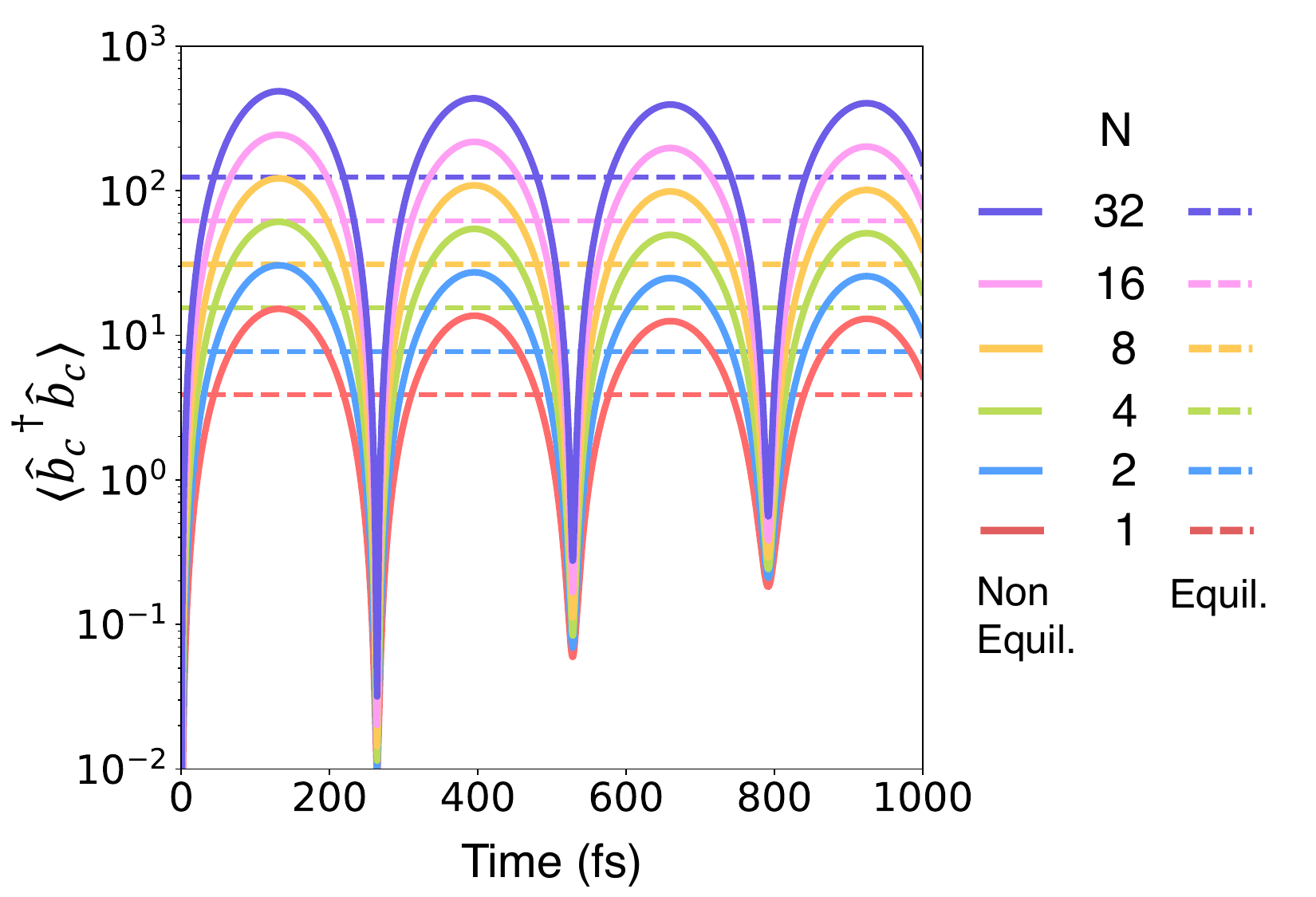}
    \caption{The time-dependent dipole-gauge cavity number operator expectation value $\langle\hat{n}_c\rangle$, for the correlated (dashed), and uncorrelated (solid lines) initial conditions for various $N$. { Results were obtained using ML-MCTDH and using the {\it constant} $\rho$ light-matter coupling Hamiltonian.}}
    \label{fig:fig7}
\end{figure}

 Finally, to explore the origins of these deviations with respect to correlated and uncorrelated initial conditions, we consider the dynamics of the cavity mode itself.  In Fig.\ref{fig:fig7}, we present the time-dependence of the dipole-gauge cavity number operator $\hat{n}_c = \hat{b}_c^\dagger \hat{b}_c$ (note that this does not correspond to the number of photons in the cavity mode).  For the case of the correlated (or equilibrated) initial conditions, $\langle\hat{n}_c\rangle$ increases linearly with the number of particles, and essesntially no dynamics are observed in $\langle\hat{n}_c\rangle$, over the timescales considered.  In contrast, for the uncorrelated initial conditions with the cavity initially prepared in the vacuum state, significant coherent oscillations are observed, consistent with the dynamics of a displaced harmonic oscillator.  For all times $t$ considered here, we observe a deviation in the expectation value from that of the correlated initial state case which scales with $N$.  As a consequence, the mean-field value for the cavity-molecule interaction observed for the correlated initial state and the vacuum initial states differ by a factor of $\sqrt{N}$.  This factor exactly cancels the $\frac{1}{\sqrt{N}}$ scaling of the cavity-molecule interaction strength and so in the $N\rightarrow\infty$ limit, the cavity modifies chemical dynamics in the collective regime when using  uncorrelated initial conditions.

\section{Summary and Conclusions}\label{sec4}
In this paper, we use exact quantum dynamics methods to look into the collective nature of cavity-modified chemical reactivity in the vibrational strong coupling regime. In particular, we investigate how a cavity can  modify chemical reactivity when coupling a set of identical molecules to cavity radiation mode in several simplified models. Specifically, we consider two model molecular systems, which we call Model I and Model II. Model I describes a set of molecules embedded in a dissipative environment which are also coupled to a cavity radiation mode. Each molecule is described by a barrier crossing dynamics that is described with a double-well potential. Model II describes a proton transfer reaction with a reaction coordinate coupled to a spectator mode. 

We simulate the cavity-modified chemical dynamics in the few-molecule limit  with our customized version of the HEOM approach (for $N=1,2$) and through the use of a ML-MCTDH based HEOM solver (for $N>2$). From the point of view of one reactive molecule, the other molecules can be seen as a dissipative bath connected to a cavity mode. The addition of several molecules renders  dissipation stronger overall, which in turn can lead to an enhancement of chemical reactivity, depending on the molecule-solvent couplings and cavity lifetime. 

To simulate the quantum dynamics of this system as $N\rightarrow \infty$, we develop a mean-field HEOM approach for addressing the thermodynamic limit. In comparison to recent work that allows for simulating polaritonic quantum dynamics for zero temperatures~\cite{JuanPNAS2023}, our mean-field HEOM approach simulates dissipative dynamics at finite temperatures in the presence of solvent degrees of freedom. Using this approach, we find the dissipative effect originating from the other molecules becomes negligible when $N\rightarrow \infty$. 
These results agree with a previous work that treated all degrees of freedom classically~\cite{MatthewJPCC2023}, as well as the results of Ref.~\cite{JuanPNAS2023}.

In this work, we also explore the importance of the choice of initial condition on the short to intermediate-time dynamics of a set of $N$ molecules interacting with a cavity mode.  We find that whether or not coupling to a cavity modifies the dynamics of an individual molecule in the limit that $N\rightarrow\infty$, depends explicitly on the initial condition.  For systems prepared in uncorrelated states, deviations between the inside cavity and out-of-cavity dynamics are observed to persist for large $N$, however, for correlated initial conditions no such deviations are observed.  

% These two initial conditions correspond to two significantly different physical situations

% \begin{itemize}
%     \item Uncorrelated: The $N$ molecules and cavity have been allowed to thermalise in the absence of the cavity and at time $t=0$ molecule-cavity interactions are introduced. 
%     \item Correlated: Interactions between the $N$ molecules and cavity have been introduced at $t\rightarrow-\infty$ and the system has been allowed to thermalize.
% \end{itemize}

% We show that the nonequilibrium decay processes, when starting with an uncorrelated initial state, arising as the system returns to equilibrium can exhibit significant modification in the collective VSC regime. However, in the case of an initially thermal configuration, we observe no such collective coupling.  

Overall, these results indicate that a non-equilibrium effects can lead to both collective and resonant modifications of chemical reactivity when coupling to cavity radiation. This finding should prompt the consideration of explicitly out of equilibrium techniques such as Floquet or pulsed laser methods, to induce non-equilibrium steady-states where altered long-time reactivity might be observed. The mean-field HEOM approach developed in this work is particularly well-suited for investigations. { In this regard, we note a recent work~\cite{ChenScience2022} which shows that VSC can alter chemistry via a nonequilibrium preparation of initial state.} On the other hand, in the context of interpreting experiments demonstrating collective VSC modification of adiabatic chemical reaction rates, our findings indicate that terms that are not incorporated in the models studied here might have relevance for the experimentally observed phenomena. Below, we list a few approximations whose role in cavity-modified chemistry should be investigated in the future. 

{\it Single mode approximation.} Most theoretical works investigating the 
VSC modification of chemistry assumes the coupling of molecules to a single cavity radiation mode~\cite{LiNC2021, LiJCP2020, TaoPNAS2020, lindoy2023quantum, SunJPCL2022, LiJPCL2021, DerekArxiv2022, MondalJCP2023,PhilbinJPCC2022}. In reality, there are an infinite set of cavity radiation modes that also have a characteristic dispersion. Recent work shows that going beyond the single-mode approximation is necessary to capture various effects in light-matter hybrid systems~\cite{HoffmannJCP2020, TichauerJCP2021, ying_taylor_huo_2023}. In the future, we will explore setups where an ensemble of cavity modes is coupled to an ensemble of molecules. 

{\it Long-wavelength approximation.} Within this approximation, the spatial variation of the radiation field is ignored. The spatial variation also plays a crucial role in the description of cavity-modified transport in materials and molecules~\cite{Xu2023, TichauerJCP2021,Sokolovskii2023}, which may potentially impact the modelling of chemical reactivity. In the near future, we will explore how the spatial variation of the radiation field impacts the chemical reactivity in the VSC regime. 

{\it Interactions between molecules.} Another approximation employed within our present simulation is that we ignore the interactions between the molecules, and they only interact via the dipole self-energy term. In previous work, it was found that cavities can modify chemical reactivities in the collective regime when coupling the rest of the molecules to a single reactive molecule~\cite{MandalJCP2022}. In contrast, here we ignore the (spatially dependent) interactions between molecules. Future work will be devoted to exploring how inter-molecular interactions may play a role beyond such extreme scenarios. It should be noted that the three aspects described above are necessary to enable and describe polariton transport in solids~\cite{Xu2023}. In addition, the spatial scale of interactions draws into question the validity of the mean-field limit, which explicitly assumes no spatial scale of interactions. One however may still employ cluster mean-field techniques~\cite{LinPRL2011, cDMFT, KochPRB2008} to attempt to build in this spatial dependence. 

Overall, our results point to the possibility of modifying chemical dynamics by coupling molecular vibrations under non-equilibrium conditions and, at the same time, indicate the inability of simple models of collective VSC to exhibit modified reactivity in the equilibrium limit. The physical  factors missing in our simple models that might allow for  a more complete understanding of the cavity-modified ground state chemical reactivity observed in in recent experiments await further investigation.

\section{Acknowledgement} 
This work was supported by NSF-2245592 (A.M. and D.R.R.) and, in the early stages of this work, by the Chemical Sciences, Geosciences, and Biosciences Division of the Office of Basic Energy Sciences, Office of Science, U.S. Department of Energy (L.P.L. and D.R.R.).  
D. R. R. acknowledges helpful discussions with Joel Yuen-Zhou and Jonathan Keeling. A. M. appreciates discussions with Pengfei Huo. 
 
\appendix
{
 \section{Numerical Validation of the Mean-Field  Population Dynamics \label{appendix:mean_field}}
In this appendix we explore the convergence of the mean-field population dynamics of model I towards the exact dynamics as $N\rightarrow\infty$.  To do this, we consider a variant of model I, in which only the four lowest energy vibrational states of the reaction coordinate are retained. This is insufficient to converge the reaction rates obtained in the main text, however, enables comparison of the population dynamics obtained using the mean-field HEOM method with exact HEOM calculations employing the ML-MCTDH solver for systems sizes up to $N=128$, for times of $t = 2$ ps. Further, we will consider the case of a cavity lifetime of $\tau_c = 10$ ps, which enables the use of a smaller hierarchy of auxiliary density operators for the cavity-loss bath.  This choice allows for calculations to be performed with up to $30$ cavity states, which are required to converge the population dynamics for large values of $N$.  
\begin{figure}[!hb]
    \centering
    \includegraphics[width=0.95\linewidth]{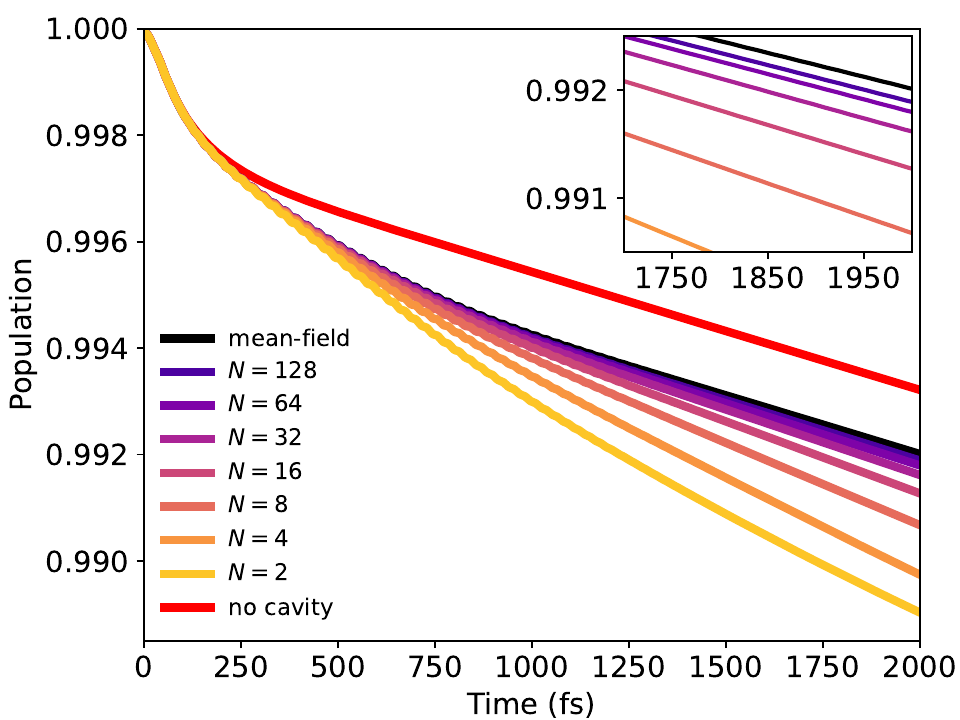}
    \caption{{ Convergence of the Mean-field cavity modified chemical dynamics in Model I under non-equilibrium (uncorrelated) initial conditions at T $=300$K. Time-dependent reactant population obtained for various values of $N$ using the numerically exact HEOM approach (with a ML-MCTDH based solver) compared to the mean-field dynamics (black solid line) obtained in the limit that $N\rightarrow\infty$ for the {\it constant} $\rho$ light-matter coupling Hamiltonian.  The cavity frequency is taken to be $\omega_c = 1185$ cm$^{-1}$. }}
    \label{fig:mean_field_comparison}
\end{figure} 
In Fig.~\ref{fig:mean_field_comparison}, we compare the mean-field population dynamics to the exact HEOM dynamics for this simplified variant of model I for various values of $N$. As was seen in Fig.~\ref{fig:fig4}, a significant deviation, from the out-of-cavity  population dynamics,  is observed in the mean-field limit.  For finite $N$, the exact HEOM calculations exhibit larger deviations than are observed in the $N\rightarrow\infty$ mean-field limit, however, these deviations are observed to decrease with increasing $N$.  These results suggest that the resonant enhancement effect observed in the few molecule regime, arises due to beyond mean-field correlations in the dynamics, that decay in the $N\rightarrow\infty$ limit.  
\\
\\
}

\bibliography{bib.bib}

\begin{thebibliography}{10}
\expandafter\ifx\csname url\endcsname\relax
  \def\url#1{\texttt{#1}}\fi
\expandafter\ifx\csname urlprefix\endcsname\relax\def\urlprefix{URL }\fi
\providecommand{\bibinfo}[2]{#2}
\providecommand{\eprint}[2][]{\url{#2}}

\bibitem{ThomasS2019}
\bibinfo{author}{Thomas, A.} \emph{et~al.}
\newblock \bibinfo{title}{Tilting a ground-state reactivity landscape by
  vibrational strong coupling}.
\newblock \emph{\bibinfo{journal}{Science}} \textbf{\bibinfo{volume}{363}},
  \bibinfo{pages}{615--619} (\bibinfo{year}{2019}).

\bibitem{ThomasACID2016}
\bibinfo{author}{Thomas, A.} \emph{et~al.}
\newblock \bibinfo{title}{Ground-state chemical reactivity under vibrational
  coupling to the vacuum electromagnetic field}.
\newblock \emph{\bibinfo{journal}{Angew. Chem. Int. Ed.}}
  \textbf{\bibinfo{volume}{128}}, \bibinfo{pages}{11634--11638}
  (\bibinfo{year}{2016}).

\bibitem{Wonmi2023S}
\bibinfo{author}{Ahn, W.}, \bibinfo{author}{Triana, J.~F.},
  \bibinfo{author}{Recabal, F.}, \bibinfo{author}{Herrera, F.} \&
  \bibinfo{author}{Simpkins, B.~S.}
\newblock \bibinfo{title}{Modification of ground-state chemical reactivity via
  light–matter coherence in infrared cavities}.
\newblock \emph{\bibinfo{journal}{Science}} \textbf{\bibinfo{volume}{380}},
  \bibinfo{pages}{1165--1168} (\bibinfo{year}{2023}).

\bibitem{NagarajanJACS2021}
\bibinfo{author}{Nagarajan, K.}, \bibinfo{author}{Thomas, A.} \&
  \bibinfo{author}{Ebbesen, T.~W.}
\newblock \bibinfo{title}{Chemistry under vibrational strong coupling}.
\newblock \emph{\bibinfo{journal}{J. Am. Chem. Soc.}}
  \textbf{\bibinfo{volume}{143}}, \bibinfo{pages}{16877--16889}
  (\bibinfo{year}{2021}).

\bibitem{MandalCR2023}
\bibinfo{author}{Mandal, A.} \emph{et~al.}
\newblock \bibinfo{title}{Theoretical advances in polariton chemistry and
  molecular cavity quantum electrodynamics}.
\newblock \emph{\bibinfo{journal}{Chem. Rev.}} \textbf{\bibinfo{volume}{123}},
  \bibinfo{pages}{9786--9879} (\bibinfo{year}{2023}).

\bibitem{LatherACID2019}
\bibinfo{author}{Lather, J.}, \bibinfo{author}{Bhatt, P.},
  \bibinfo{author}{Thomas, A.}, \bibinfo{author}{Ebbesen, T.~W.} \&
  \bibinfo{author}{George, J.}
\newblock \bibinfo{title}{Cavity catalysis by cooperative vibrational strong
  coupling of reactant and solvent molecules}.
\newblock \emph{\bibinfo{journal}{Angew. Chem. Int. Ed.}}
  \bibinfo{pages}{10635--10638} (\bibinfo{year}{2019}).

\bibitem{LatherCS2022}
\bibinfo{author}{Lather, J.}, \bibinfo{author}{Thabassum, A. N.~K.},
  \bibinfo{author}{Singh, J.} \& \bibinfo{author}{George, J.}
\newblock \bibinfo{title}{Cavity catalysis: Modifying linear free-energy
  relationship under cooperative vibrational strong coupling}.
\newblock \emph{\bibinfo{journal}{Chem. Sci.}} \textbf{\bibinfo{volume}{13}},
  \bibinfo{pages}{195--202} (\bibinfo{year}{2022}).

\bibitem{AnoopNp2020}
\bibinfo{author}{Thomas, A.} \emph{et~al.}
\newblock \bibinfo{title}{Ground state chemistry under vibrational strong
  coupling: Dependence of thermodynamic parameters on the rabi splitting
  energy}.
\newblock \emph{\bibinfo{journal}{Nanophotonics}} \textbf{\bibinfo{volume}{9}},
  \bibinfo{pages}{249--255} (\bibinfo{year}{2020}).

\bibitem{LindoyJPCL2022}
\bibinfo{author}{Lindoy, L.~P.}, \bibinfo{author}{Mandal, A.} \&
  \bibinfo{author}{Reichman, D.~R.}
\newblock \bibinfo{title}{Resonant cavity modification of ground-state chemical
  kinetics}.
\newblock \emph{\bibinfo{journal}{J. Phys. Chem. Lett.}}
  \textbf{\bibinfo{volume}{13}}, \bibinfo{pages}{6580--6586}
  (\bibinfo{year}{2022}).

\bibitem{lindoy2023quantum}
\bibinfo{author}{Lindoy, L.~P.}, \bibinfo{author}{Mandal, A.} \&
  \bibinfo{author}{Reichman, D.~R.}
\newblock \bibinfo{title}{Quantum dynamical effects of vibrational strong
  coupling in chemical reactivity}.
\newblock \emph{\bibinfo{journal}{Nat. Commun.}} \textbf{\bibinfo{volume}{14}},
  \bibinfo{pages}{2733} (\bibinfo{year}{2023}).

\bibitem{schafer2022shining}
\bibinfo{author}{Sch{\"a}fer, C.}, \bibinfo{author}{Flick, J.},
  \bibinfo{author}{Ronca, E.}, \bibinfo{author}{Narang, P.} \&
  \bibinfo{author}{Rubio, A.}
\newblock \bibinfo{title}{Shining light on the microscopic resonant mechanism
  responsible for cavity-mediated chemical reactivity}.
\newblock \emph{\bibinfo{journal}{Nat. Commun.}} \textbf{\bibinfo{volume}{13}},
  \bibinfo{pages}{7817} (\bibinfo{year}{2022}).

\bibitem{LiNC2021}
\bibinfo{author}{Li, X.}, \bibinfo{author}{Mandal, A.} \& \bibinfo{author}{Huo,
  P.}
\newblock \bibinfo{title}{Cavity frequency-dependent theory for vibrational
  polariton chemistry}.
\newblock \emph{\bibinfo{journal}{Nat. Commun.}} \textbf{\bibinfo{volume}{12}},
  \bibinfo{pages}{1315} (\bibinfo{year}{2021}).

\bibitem{LiJPCL2021}
\bibinfo{author}{Li, X.}, \bibinfo{author}{Mandal, A.} \& \bibinfo{author}{Huo,
  P.}
\newblock \bibinfo{title}{Theory of mode-selective chemistry through
  polaritonic vibrational strong coupling}.
\newblock \emph{\bibinfo{journal}{J. Phys. Chem. Lett.}}
  \textbf{\bibinfo{volume}{12}}, \bibinfo{pages}{6974--6982}
  (\bibinfo{year}{2021}).

\bibitem{MandalJCP2022}
\bibinfo{author}{Mandal, A.}, \bibinfo{author}{Li, X.} \& \bibinfo{author}{Huo,
  P.}
\newblock \bibinfo{title}{Theory of vibrational polariton chemistry in the
  collective coupling regime}.
\newblock \emph{\bibinfo{journal}{J. Chem. Phys.}}
  \textbf{\bibinfo{volume}{156}}, \bibinfo{pages}{014101}
  (\bibinfo{year}{2022}).

\bibitem{MatthewJPCC2023}
\bibinfo{author}{Du, M.}, \bibinfo{author}{Poh, Y.~R.} \&
  \bibinfo{author}{Yuen-Zhou, J.}
\newblock \bibinfo{title}{Vibropolaritonic reaction rates in the collective
  strong coupling regime: Pollak–grabert–hänggi theory}.
\newblock \emph{\bibinfo{journal}{J. Phys. Chem. C}}
  \textbf{\bibinfo{volume}{127}}, \bibinfo{pages}{5230--5237}
  (\bibinfo{year}{2023}).

\bibitem{campos2023swinging}
\bibinfo{author}{Campos-Gonzalez-Angulo, J.}, \bibinfo{author}{Poh, Y.},
  \bibinfo{author}{Du, M.} \& \bibinfo{author}{Yuen-Zhou, J.}
\newblock \bibinfo{title}{Swinging between shine and shadow: Theoretical
  advances on thermally activated vibropolaritonic chemistry}.
\newblock \emph{\bibinfo{journal}{J. Chem. Phys.}}
  \textbf{\bibinfo{volume}{158}} (\bibinfo{year}{2023}).

\bibitem{mondal2022dissociation}
\bibinfo{author}{Mondal, S.}, \bibinfo{author}{Wang, D.~S.} \&
  \bibinfo{author}{Keshavamurthy, S.}
\newblock \bibinfo{title}{Dissociation dynamics of a diatomic molecule in an
  optical cavity}.
\newblock \emph{\bibinfo{journal}{J. Chem. Phys.}}
  \textbf{\bibinfo{volume}{157}} (\bibinfo{year}{2022}).

\bibitem{sun2023modification}
\bibinfo{author}{Sun, J.} \& \bibinfo{author}{Vendrell, O.}
\newblock \bibinfo{title}{Modification of thermal chemical rates in a cavity
  via resonant effects in the collective regime}.
\newblock \emph{\bibinfo{journal}{J. Phys. Chem. Lett.}}
  \textbf{\bibinfo{volume}{14}}, \bibinfo{pages}{8397--8404}
  (\bibinfo{year}{2023}).

\bibitem{fischer2023cavity}
\bibinfo{author}{Fischer, E.~W.} \& \bibinfo{author}{Saalfrank, P.}
\newblock \bibinfo{title}{Cavity-catalyzed hydrogen transfer dynamics in an
  entangled molecular ensemble under vibrational strong coupling}.
\newblock \emph{\bibinfo{journal}{Phys. Chem. Chem. Phys.}}
  \textbf{\bibinfo{volume}{25}}, \bibinfo{pages}{11771--11779}
  (\bibinfo{year}{2023}).

\bibitem{fischer2023beyond}
\bibinfo{author}{Fischer, E.~W.} \& \bibinfo{author}{Saalfrank, P.}
\newblock \bibinfo{title}{Beyond cavity born--oppenheimer: On nonadiabatic
  coupling and effective ground state hamiltonians in vibro-polaritonic
  chemistry}.
\newblock \emph{\bibinfo{journal}{J. Chem. Theory Comput.}}
  (\bibinfo{year}{2023}).

\bibitem{YingJCP2023}
\bibinfo{author}{Ying, W.} \& \bibinfo{author}{Huo, P.}
\newblock \bibinfo{title}{{Resonance theory and quantum dynamics simulations of
  vibrational polariton chemistry}}.
\newblock \emph{\bibinfo{journal}{J. Chem. Phys.}}
  \textbf{\bibinfo{volume}{159}}, \bibinfo{pages}{084104}
  (\bibinfo{year}{2023}).

\bibitem{PhilbinJPCC2022}
\bibinfo{author}{Philbin, J.~P.}, \bibinfo{author}{Wang, Y.},
  \bibinfo{author}{Narang, P.} \& \bibinfo{author}{Dou, W.}
\newblock \bibinfo{title}{Chemical reactions in imperfect cavities:
  Enhancement, suppression, and resonance}.
\newblock \emph{\bibinfo{journal}{J. Phys. Chem. C}}
  \textbf{\bibinfo{volume}{126}}, \bibinfo{pages}{14908--14913}
  (\bibinfo{year}{2022}).

\bibitem{FiechterJCPL2023}
\bibinfo{author}{Fiechter, M.~R.}, \bibinfo{author}{Runeson, J.~E.},
  \bibinfo{author}{Lawrence, J.~E.} \& \bibinfo{author}{Richardson, J.~O.}
\newblock \bibinfo{title}{How quantum is the resonance behavior in vibrational
  polariton chemistry?}
\newblock \emph{\bibinfo{journal}{J. Phys. Chem. Lett.}}
  \textbf{\bibinfo{volume}{14}}, \bibinfo{pages}{8261--8267}
  (\bibinfo{year}{2023}).

\bibitem{ImperatoreJCP2021}
\bibinfo{author}{Imperatore, M.~V.}, \bibinfo{author}{Asbury, J.~B.} \&
  \bibinfo{author}{Giebink, N.~C.}
\newblock \bibinfo{title}{Reproducibility of cavity-enhanced chemical reaction
  rates in the vibrational strong coupling regime}.
\newblock \emph{\bibinfo{journal}{J. Chem. Phys.}}
  \textbf{\bibinfo{volume}{154}}, \bibinfo{pages}{191103}
  (\bibinfo{year}{2021}).

\bibitem{WiesehanJCP2021}
\bibinfo{author}{Wiesehan, G.~D.} \& \bibinfo{author}{Xiong, W.}
\newblock \bibinfo{title}{Negligible rate enhancement from reported cooperative
  vibrational strong coupling catalysis}.
\newblock \emph{\bibinfo{journal}{J. Chem. Phys.}}
  \textbf{\bibinfo{volume}{155}}, \bibinfo{pages}{241103}
  (\bibinfo{year}{2021}).

\bibitem{MondalJPCL2022}
\bibinfo{author}{Mondal, M.}, \bibinfo{author}{Semenov, A.},
  \bibinfo{author}{Ochoa, M.~A.} \& \bibinfo{author}{Nitzan, A.}
\newblock \bibinfo{title}{Strong coupling in infrared plasmonic cavities}.
\newblock \emph{\bibinfo{journal}{J. Phys. Chem. Lett.}}
  \textbf{\bibinfo{volume}{13}}, \bibinfo{pages}{9673--9678}
  (\bibinfo{year}{2022}).

\bibitem{DerekArxiv2022}
\bibinfo{author}{Wang, D.~S.}, \bibinfo{author}{Flick, J.} \&
  \bibinfo{author}{Yelin, S.~F.}
\newblock \bibinfo{title}{Chemical reactivity under collective vibrational
  strong coupling}.
\newblock \emph{\bibinfo{journal}{J. Chem. Phys.}}
  \textbf{\bibinfo{volume}{157}}, \bibinfo{pages}{224304}
  (\bibinfo{year}{2022}).

\bibitem{PhilbinJPCL2023}
\bibinfo{author}{Philbin, J.~P.} \emph{et~al.}
\newblock \bibinfo{title}{Molecular van der waals fluids in cavity quantum
  electrodynamics}.
\newblock \emph{\bibinfo{journal}{J Phys. Chem. Lett}}
  \textbf{\bibinfo{volume}{14}}, \bibinfo{pages}{8988--8993}
  (\bibinfo{year}{2023}).

\bibitem{JuanPNAS2023}
\bibinfo{author}{Pérez-Sánchez, J.~B.}, \bibinfo{author}{Koner, A.},
  \bibinfo{author}{Stern, N.~P.} \& \bibinfo{author}{Yuen-Zhou, J.}
\newblock \bibinfo{title}{Simulating molecular polaritons in the collective
  regime using few-molecule models}.
\newblock \emph{\bibinfo{journal}{Proc. Natl. Acad. Sci. U.S.A.}}
  \textbf{\bibinfo{volume}{120}}, \bibinfo{pages}{e2219223120}
  (\bibinfo{year}{2023}).

\bibitem{CALDEIRA1983374}
\bibinfo{author}{Caldeira, A.} \& \bibinfo{author}{Leggett, A.}
\newblock \bibinfo{title}{Quantum tunnelling in a dissipative system}.
\newblock \emph{\bibinfo{journal}{Ann. Phys.}} \textbf{\bibinfo{volume}{149}},
  \bibinfo{pages}{374--456} (\bibinfo{year}{1983}).

\bibitem{keeling_notes}
\bibinfo{author}{Keeling, J.}
\newblock \bibinfo{title}{Light-matter interactions and quantum optics}.
\newblock
  \urlprefix\url{https://www.st-andrews.ac.uk/~jmjk/keeling/teaching/quantum-optics.pdf}.

\bibitem{Thirunamachandran1998}
\bibinfo{author}{Craig, D.~P.} \& \bibinfo{author}{Thirunamachandran, T.}
\newblock \emph{\bibinfo{title}{Molecular Quantum Electrodynamics: An
  Introduction to Radiation-Molecule Interactions}} (\bibinfo{publisher}{Dover
  Publications}, \bibinfo{year}{1998}).

\bibitem{Bernardis2018PRA}
\bibinfo{author}{Bernardis, D.~D.}, \bibinfo{author}{Jaako, T.} \&
  \bibinfo{author}{Rabl, P.}
\newblock \bibinfo{title}{Cavity quantum electrodynamics in the nonperturbative
  regime}.
\newblock \emph{\bibinfo{journal}{Phys. Rev. A}} \textbf{\bibinfo{volume}{97}},
  \bibinfo{pages}{043820} (\bibinfo{year}{2018}).

\bibitem{MandalNL2023}
\bibinfo{author}{Mandal, A.} \emph{et~al.}
\newblock \bibinfo{title}{Microscopic theory of multimode polariton dispersion
  in multilayered materials}.
\newblock \emph{\bibinfo{journal}{Nano Lett.}} \textbf{\bibinfo{volume}{23}},
  \bibinfo{pages}{4082--4089} (\bibinfo{year}{2023}).

\bibitem{TaoACIE2021}
\bibinfo{author}{Li, T.~E.}, \bibinfo{author}{Nitzan, A.} \&
  \bibinfo{author}{Subotnik, J.~E.}
\newblock \bibinfo{title}{Collective vibrational strong coupling effects on
  molecular vibrational relaxation and energy transfer: Numerical insights via
  cavity molecular dynamics simulations**}.
\newblock \emph{\bibinfo{journal}{Angew. Chem. Int. Ed.}}
  \textbf{\bibinfo{volume}{60}}, \bibinfo{pages}{15533--15540}
  (\bibinfo{year}{2021}).

\bibitem{TaoPNAS2020}
\bibinfo{author}{Li, T.~E.}, \bibinfo{author}{Subotnik, J.~E.} \&
  \bibinfo{author}{Nitzan, A.}
\newblock \bibinfo{title}{Cavity molecular dynamics simulations of liquid water
  under vibrational ultrastrong coupling}.
\newblock \emph{\bibinfo{journal}{Proc. Natl. Acad. Sci. U.S.A.}}
  \textbf{\bibinfo{volume}{117}}, \bibinfo{pages}{18324--18331}
  (\bibinfo{year}{2020}).

\bibitem{YechemArxiv2023}
\bibinfo{author}{Ke, Y.} \& \bibinfo{author}{Richardson, J.~O.}
\newblock \bibinfo{title}{Insights into the mechanisms of optical
  cavity-modified ground-state}.
\newblock \emph{\bibinfo{journal}{ChemRxiv}}  (\bibinfo{year}{2023}).

\bibitem{GustinPNAS2023}
\bibinfo{author}{Gustin, I.}, \bibinfo{author}{Kim, C.~W.},
  \bibinfo{author}{McCamant, D.~W.} \& \bibinfo{author}{Franco, I.}
\newblock \bibinfo{title}{Mapping electronic decoherence pathways in
  molecules}.
\newblock \emph{\bibinfo{journal}{Proceedings of the National Academy of
  Sciences}} \textbf{\bibinfo{volume}{120}}, \bibinfo{pages}{e2309987120}
  (\bibinfo{year}{2023}).

\bibitem{UenoJCTC2020}
\bibinfo{author}{Ueno, S.} \& \bibinfo{author}{Tanimura, Y.}
\newblock \bibinfo{title}{Modeling intermolecular and intramolecular modes of
  liquid water using multiple heat baths: Machine learning approach}.
\newblock \emph{\bibinfo{journal}{Journal of Chemical Theory and Computation}}
  \textbf{\bibinfo{volume}{16}}, \bibinfo{pages}{2099--2108}
  (\bibinfo{year}{2020}).

\bibitem{SakuraiJPCA2011}
\bibinfo{author}{Sakurai, A.} \& \bibinfo{author}{Tanimura, Y.}
\newblock \bibinfo{title}{Does $\hbar$ play a role in multidimensional
  spectroscopy? reduced hierarchy equations of motion approach to molecular
  vibrations}.
\newblock \emph{\bibinfo{journal}{The Journal of Physical Chemistry A}}
  \textbf{\bibinfo{volume}{115}}, \bibinfo{pages}{4009--4022}
  (\bibinfo{year}{2011}).

\bibitem{DoslicPCCP1999}
\bibinfo{author}{Doslic, N.} \emph{et~al.}
\newblock \bibinfo{title}{Ultrafast photoinduced dissipative hydrogen switching
  dynamics in thioacetylacetone}.
\newblock \emph{\bibinfo{journal}{Phys. Chem. Chem. Phys.}}
  \textbf{\bibinfo{volume}{1}}, \bibinfo{pages}{1249--1257}
  (\bibinfo{year}{1999}).

\bibitem{10.1063/1.462100}
\bibinfo{author}{Colbert, D.~T.} \& \bibinfo{author}{Miller, W.~H.}
\newblock \bibinfo{title}{{A novel discrete variable representation for quantum
  mechanical reactive scattering via the S‐matrix Kohn method}}.
\newblock \emph{\bibinfo{journal}{J. Chem. Phys.}}
  \textbf{\bibinfo{volume}{96}}, \bibinfo{pages}{1982--1991}
  (\bibinfo{year}{1992}).

\bibitem{TanimuraJPSP1989}
\bibinfo{author}{Tanimura, Y.} \& \bibinfo{author}{Kubo, R.}
\newblock \bibinfo{title}{Time evolution of a quantum system in contact with a
  nearly gaussian-markoffian noise bath}.
\newblock \emph{\bibinfo{journal}{J. Phys. Soc. Jap.}}
  \textbf{\bibinfo{volume}{58}}, \bibinfo{pages}{101--114}
  (\bibinfo{year}{1989}).

\bibitem{IshizakiJPSJ2005}
\bibinfo{author}{Ishizaki, A.} \& \bibinfo{author}{Tanimura, Y.}
\newblock \bibinfo{title}{Quantum dynamics of system strongly coupled to
  low-temperature colored noise bath: Reduced hierarchy equations approach}.
\newblock \emph{\bibinfo{journal}{J. Phys. Soc. Jpn.}}
  \textbf{\bibinfo{volume}{74}}, \bibinfo{pages}{3131--3134}
  (\bibinfo{year}{2005}).

\bibitem{TanimuraJCP2020}
\bibinfo{author}{Tanimura, Y.}
\newblock \bibinfo{title}{Numerically “exact” approach to open quantum
  dynamics: The hierarchical equations of motion (heom)}.
\newblock \emph{\bibinfo{journal}{J. Chem. Phys.}}
  \textbf{\bibinfo{volume}{153}}, \bibinfo{pages}{020901}
  (\bibinfo{year}{2020}).

\bibitem{lindoythesis}
\bibinfo{author}{Lindoy, L. P.~J.}
\newblock \emph{\bibinfo{title}{New Developments in Open System Quantum
  Dynamics}}.
\newblock \bibinfo{type}{D.phil. thesis}, \bibinfo{school}{Magdalen College,
  University of Oxford} (\bibinfo{year}{2019}).

\bibitem{10.1063/5.0050720}
\bibinfo{author}{Yan, Y.}, \bibinfo{author}{Xu, M.}, \bibinfo{author}{Li, T.}
  \& \bibinfo{author}{Shi, Q.}
\newblock \bibinfo{title}{{Efficient propagation of the hierarchical equations
  of motion using the Tucker and hierarchical Tucker tensors}}.
\newblock \emph{\bibinfo{journal}{J. Chem. Phys.}}
  \textbf{\bibinfo{volume}{154}}, \bibinfo{pages}{194104}
  (\bibinfo{year}{2021}).

\bibitem{10.1063/5.0153870}
\bibinfo{author}{Ke, Y.}
\newblock \bibinfo{title}{{Tree tensor network state approach for solving
  hierarchical equations of motion}}.
\newblock \emph{\bibinfo{journal}{J. Chem. Phys.}}
  \textbf{\bibinfo{volume}{158}}, \bibinfo{pages}{211102}
  (\bibinfo{year}{2023}).

\bibitem{PiperPRL2022}
\bibinfo{author}{Fowler-Wright, P.}, \bibinfo{author}{Lovett, B.~W.} \&
  \bibinfo{author}{Keeling, J.}
\newblock \bibinfo{title}{Efficient many-body non-markovian dynamics of organic
  polaritons}.
\newblock \emph{\bibinfo{journal}{Phys. Rev. Lett.}}
  \textbf{\bibinfo{volume}{129}}, \bibinfo{pages}{173001}
  (\bibinfo{year}{2022}).

\bibitem{Mori_2013}
\bibinfo{author}{Mori, T.}
\newblock \bibinfo{title}{Exactness of the mean-field dynamics in optical
  cavity systems}.
\newblock \emph{\bibinfo{journal}{J. Stat. Mech.: Theory Exp}}
  \textbf{\bibinfo{volume}{2013}}, \bibinfo{pages}{P06005}
  (\bibinfo{year}{2013}).

\bibitem{CarolloPRL2021}
\bibinfo{author}{Carollo, F.} \& \bibinfo{author}{Lesanovsky, I.}
\newblock \bibinfo{title}{Exactness of mean-field equations for open dicke
  models with an application to pattern retrieval dynamics}.
\newblock \emph{\bibinfo{journal}{Phys. Rev. Lett.}}
  \textbf{\bibinfo{volume}{126}}, \bibinfo{pages}{230601}
  (\bibinfo{year}{2021}).

\bibitem{doi:10.1063/1.5116800}
\bibinfo{author}{Lawrence, J.~E.}, \bibinfo{author}{Fletcher, T.},
  \bibinfo{author}{Lindoy, L.~P.} \& \bibinfo{author}{Manolopoulos, D.~E.}
\newblock \bibinfo{title}{On the calculation of quantum mechanical electron
  transfer rates}.
\newblock \emph{\bibinfo{journal}{J. Chem. Phys.}}
  \textbf{\bibinfo{volume}{151}}, \bibinfo{pages}{114119}
  (\bibinfo{year}{2019}).

\bibitem{doi:10.1063/5.0098545}
\bibinfo{author}{Ke, Y.}, \bibinfo{author}{Kaspar, C.},
  \bibinfo{author}{Erpenbeck, A.}, \bibinfo{author}{Peskin, U.} \&
  \bibinfo{author}{Thoss, M.}
\newblock \bibinfo{title}{Nonequilibrium reaction rate theory: Formulation and
  implementation within the hierarchical equations of motion approach}.
\newblock \emph{\bibinfo{journal}{J. Chem. Phys.}}
  \textbf{\bibinfo{volume}{157}}, \bibinfo{pages}{034103}
  (\bibinfo{year}{2022}).

\bibitem{QiangJCP2011}
\bibinfo{author}{Shi, Q.}, \bibinfo{author}{Zhu, L.} \& \bibinfo{author}{Chen,
  L.}
\newblock \bibinfo{title}{Quantum rate dynamics for proton transfer reaction in
  a model system: Effect of the rate promoting vibrational mode}.
\newblock \emph{\bibinfo{journal}{J. Chem. Phys.}}
  \textbf{\bibinfo{volume}{135}}, \bibinfo{pages}{044505}
  (\bibinfo{year}{2011}).

\bibitem{doi:10.1063/1.4890441}
\bibinfo{author}{Tanimura, Y.}
\newblock \bibinfo{title}{Reduced hierarchical equations of motion in real and
  imaginary time: Correlated initial states and thermodynamic quantities}.
\newblock \emph{\bibinfo{journal}{J. Chem. Phys.}}
  \textbf{\bibinfo{volume}{141}}, \bibinfo{pages}{044114}
  (\bibinfo{year}{2014}).

\bibitem{doi:10.1063/1.2772265}
\bibinfo{author}{Craig, I.~R.}, \bibinfo{author}{Thoss, M.} \&
  \bibinfo{author}{Wang, H.}
\newblock \bibinfo{title}{Proton transfer reactions in model condensed-phase
  environments: Accurate quantum dynamics using the multilayer
  multiconfiguration time-dependent hartree approach}.
\newblock \emph{\bibinfo{journal}{J. Chem. Phys.}}
  \textbf{\bibinfo{volume}{127}}, \bibinfo{pages}{144503}
  (\bibinfo{year}{2007}).

\bibitem{LUBICH2015}
\bibinfo{author}{Lubich, C.}
\newblock \bibinfo{title}{{Time Integration in the Multiconfiguration
  Time-Dependent Hartree Method of Molecular Quantum Dynamics}}.
\newblock \emph{\bibinfo{journal}{Appl. Math. Res. Express}}
  \textbf{\bibinfo{volume}{2015}}, \bibinfo{pages}{311--328}
  (\bibinfo{year}{2015}).

\bibitem{KieriSJNA2016}
\bibinfo{author}{Kieri, E.}, \bibinfo{author}{Lubich, C.} \&
  \bibinfo{author}{Walach, H.}
\newblock \bibinfo{title}{Discretized dynamical low-rank approximation in the
  presence of small singular values}.
\newblock \emph{\bibinfo{journal}{SIAM J. Numer. Anal.}}
  \textbf{\bibinfo{volume}{54}}, \bibinfo{pages}{1020--1038}
  (\bibinfo{year}{2016}).

\bibitem{KLOSS2017}
\bibinfo{author}{Kloss, B.}, \bibinfo{author}{Burghardt, I.} \&
  \bibinfo{author}{Lubich, C.}
\newblock \bibinfo{title}{Implementation of a novel projector-splitting
  integrator for the multi-configurational time-dependent hartree approach}.
\newblock \emph{\bibinfo{journal}{J. Chem. Phys.}}
  \textbf{\bibinfo{volume}{146}}, \bibinfo{pages}{174107}
  (\bibinfo{year}{2017}).

\bibitem{BONFANTI2018252}
\bibinfo{author}{Bonfanti, M.} \& \bibinfo{author}{Burghardt, I.}
\newblock \bibinfo{title}{Tangent space formulation of the multi-configuration
  time-dependent hartree equations of motion: The projector-splitting algorithm
  revisited}.
\newblock \emph{\bibinfo{journal}{Chem. Phys.}} \textbf{\bibinfo{volume}{515}},
  \bibinfo{pages}{252--261} (\bibinfo{year}{2018}).

\bibitem{CERUTI2021}
\bibinfo{author}{Ceruti, G.}, \bibinfo{author}{Lubich, C.} \&
  \bibinfo{author}{Walach, H.}
\newblock \bibinfo{title}{Time integration of tree tensor networks}.
\newblock \emph{\bibinfo{journal}{SIAM J. Numer. Anal.}}
  \textbf{\bibinfo{volume}{59}}, \bibinfo{pages}{289--313}
  (\bibinfo{year}{2021}).

\bibitem{lindoy_mctdh_1}
\bibinfo{author}{Lindoy, L.~P.}, \bibinfo{author}{Kloss, B.} \&
  \bibinfo{author}{Reichman, D.~R.}
\newblock \bibinfo{title}{Time evolution of ml-mctdh wavefunctions. i. gauge
  conditions, basis functions, and singularities}.
\newblock \emph{\bibinfo{journal}{J. Chem. Phys.}}
  \textbf{\bibinfo{volume}{155}}, \bibinfo{pages}{174108}
  (\bibinfo{year}{2021}).

\bibitem{lindoy_mctdh_2}
\bibinfo{author}{Lindoy, L.~P.}, \bibinfo{author}{Kloss, B.} \&
  \bibinfo{author}{Reichman, D.~R.}
\newblock \bibinfo{title}{Time evolution of ml-mctdh wavefunctions. ii.
  application of the projector splitting integrator}.
\newblock \emph{\bibinfo{journal}{J. Chem. Phys.}}
  \textbf{\bibinfo{volume}{155}}, \bibinfo{pages}{174109}
  (\bibinfo{year}{2021}).

\bibitem{MendiveJCP2020}
\bibinfo{author}{Mendive-Tapia, D.} \& \bibinfo{author}{Meyer, H.-D.}
\newblock \bibinfo{title}{{Regularizing the MCTDH equations of motion through
  an optimal choice on-the-fly (i.e., spawning) of unoccupied single-particle
  functions}}.
\newblock \emph{\bibinfo{journal}{J. Chem. Phys.}}
  \textbf{\bibinfo{volume}{153}}, \bibinfo{pages}{234114}
  (\bibinfo{year}{2020}).

\bibitem{LarssonJCP2019}
\bibinfo{author}{Larsson, H.~R.}
\newblock \bibinfo{title}{Computing vibrational eigenstates with tree tensor
  network states (ttns)}.
\newblock \emph{\bibinfo{journal}{J. Chem. Phys.}}
  \textbf{\bibinfo{volume}{151}}, \bibinfo{pages}{204102}
  (\bibinfo{year}{2019}).

\bibitem{Llindoy}
\bibinfo{author}{Llindoy}.
\newblock \bibinfo{title}{Llindoy/ttns$\_$lib}.
\newblock \urlprefix\url{https://github.com/llindoy/ttns_lib}.

\bibitem{FeistNP2021}
\bibinfo{author}{Feist, J.}, \bibinfo{author}{Fernández-Domínguez, A.~I.} \&
  \bibinfo{author}{García-Vidal, F.~J.}
\newblock \bibinfo{title}{Macroscopic qed for quantum nanophotonics:
  emitter-centered modes as a minimal basis for multiemitter problems}.
\newblock \emph{\bibinfo{journal}{Nanophotonics}}
  \textbf{\bibinfo{volume}{10}}, \bibinfo{pages}{477--489}
  (\bibinfo{year}{2021}).

\bibitem{Christian2020}
\bibinfo{author}{Schäfer, C.}, \bibinfo{author}{Ruggenthaler, M.},
  \bibinfo{author}{Rokaj, V.} \& \bibinfo{author}{Rubio, A.}
\newblock \bibinfo{title}{Relevance of the quadratic diamagnetic and
  self-polarization terms in cavity quantum electrodynamics}.
\newblock \emph{\bibinfo{journal}{ACS Photonics}} \textbf{\bibinfo{volume}{7}},
  \bibinfo{pages}{975--990} (\bibinfo{year}{2020}).

\bibitem{TaylorOL2022}
\bibinfo{author}{Taylor, M. A.~D.}, \bibinfo{author}{Mandal, A.} \&
  \bibinfo{author}{Huo, P.}
\newblock \bibinfo{title}{Resolving ambiguities of the mode truncation in
  cavity quantum electrodynamics}.
\newblock \emph{\bibinfo{journal}{Opt. Lett.}} \textbf{\bibinfo{volume}{47}},
  \bibinfo{pages}{1446--1449} (\bibinfo{year}{2022}).

\bibitem{RokajJPB2018}
\bibinfo{author}{Rokaj, V.}, \bibinfo{author}{Welakuh, D.~M.},
  \bibinfo{author}{Ruggenthaler, M.} \& \bibinfo{author}{Rubio, A.}
\newblock \bibinfo{title}{Light–matter interaction in the long-wavelength
  limit: no ground-state without dipole self-energy}.
\newblock \emph{\bibinfo{journal}{J. Phys. B: At. Mol. Opt. Phys.}}
  \textbf{\bibinfo{volume}{51}}, \bibinfo{pages}{034005}
  (\bibinfo{year}{2018}).

\bibitem{GalegoPRX2019}
\bibinfo{author}{Galego, J.}, \bibinfo{author}{Climent, C.},
  \bibinfo{author}{Garcia-Vidal, F.~J.} \& \bibinfo{author}{Feist, J.}
\newblock \bibinfo{title}{Cavity casimir-polder forces and their effects in
  ground-state chemical reactivity}.
\newblock \emph{\bibinfo{journal}{Phys. Rev. X}} \textbf{\bibinfo{volume}{9}},
  \bibinfo{pages}{021057} (\bibinfo{year}{2019}).

\bibitem{MichettiPRB2005}
\bibinfo{author}{Michetti, P.} \& \bibinfo{author}{La~Rocca, G.~C.}
\newblock \bibinfo{title}{Polariton states in disordered organic
  microcavities}.
\newblock \emph{\bibinfo{journal}{Phys. Rev. B}} \textbf{\bibinfo{volume}{71}},
  \bibinfo{pages}{115320} (\bibinfo{year}{2005}).

\bibitem{SuyabatmazJCP2023}
\bibinfo{author}{Suyabatmaz, E.} \& \bibinfo{author}{Ribeiro, R.~F.}
\newblock \bibinfo{title}{{Vibrational polariton transport in disordered
  media}}.
\newblock \emph{\bibinfo{journal}{The Journal of Chemical Physics}}
  \textbf{\bibinfo{volume}{159}}, \bibinfo{pages}{034701}
  (\bibinfo{year}{2023}).
\newblock \urlprefix\url{https://doi.org/10.1063/5.0156008}.

\bibitem{AroeiraNP2024}
\bibinfo{author}{Aroeira, G. J.~R.}, \bibinfo{author}{Kairys, K.~T.} \&
  \bibinfo{author}{Ribeiro, R.~F.}
\newblock \bibinfo{title}{Coherent transient exciton transport in disordered
  polaritonic wires}.
\newblock \emph{\bibinfo{journal}{Nanophotonics}}  (\bibinfo{year}{2024}).

\bibitem{berghuis2022controlling}
\bibinfo{author}{Berghuis, A.~M.} \emph{et~al.}
\newblock \bibinfo{title}{Controlling exciton propagation in organic crystals
  through strong coupling to plasmonic nanoparticle arrays}.
\newblock \emph{\bibinfo{journal}{ACS photonics}} \textbf{\bibinfo{volume}{9}},
  \bibinfo{pages}{2263--2272} (\bibinfo{year}{2022}).

\bibitem{Xu2023}
\bibinfo{author}{Xu, D.} \emph{et~al.}
\newblock \bibinfo{title}{Ultrafast imaging of polariton propagation and
  interactions}.
\newblock \emph{\bibinfo{journal}{Nat. Commun.}} \textbf{\bibinfo{volume}{14}},
  \bibinfo{pages}{3881} (\bibinfo{year}{2023}).

\bibitem{QiuJPCL2021}
\bibinfo{author}{Qiu, L.} \emph{et~al.}
\newblock \bibinfo{title}{Molecular polaritons generated from strong coupling
  between cdse nanoplatelets and a dielectric optical cavity}.
\newblock \emph{\bibinfo{journal}{The Journal of Physical Chemistry Letters}}
  \textbf{\bibinfo{volume}{12}}, \bibinfo{pages}{5030--5038}
  (\bibinfo{year}{2021}).

\bibitem{EngelhardtPRL2023}
\bibinfo{author}{Engelhardt, G.} \& \bibinfo{author}{Cao, J.}
\newblock \bibinfo{title}{Polariton localization and dispersion properties of
  disordered quantum emitters in multimode microcavities}.
\newblock \emph{\bibinfo{journal}{Phys. Rev. Lett.}}
  \textbf{\bibinfo{volume}{130}}, \bibinfo{pages}{213602}
  (\bibinfo{year}{2023}).

\bibitem{DuPRL2022}
\bibinfo{author}{Du, M.} \& \bibinfo{author}{Yuen-Zhou, J.}
\newblock \bibinfo{title}{Catalysis by dark states in vibropolaritonic
  chemistry}.
\newblock \emph{\bibinfo{journal}{Phys. Rev. Lett.}}
  \textbf{\bibinfo{volume}{128}}, \bibinfo{pages}{096001}
  (\bibinfo{year}{2022}).

\bibitem{SunJPCL2022}
\bibinfo{author}{Sun, J.} \& \bibinfo{author}{Vendrell, O.}
\newblock \bibinfo{title}{Suppression and enhancement of thermal chemical rates
  in a cavity}.
\newblock \emph{\bibinfo{journal}{J. Phys. Chem. Lett.}}
  \textbf{\bibinfo{volume}{13}}, \bibinfo{pages}{4441--4446}
  (\bibinfo{year}{2022}).

\bibitem{ChenScience2022}
\bibinfo{author}{Chen, T.-T.}, \bibinfo{author}{Du, M.}, \bibinfo{author}{Yang,
  Z.}, \bibinfo{author}{Yuen-Zhou, J.} \& \bibinfo{author}{Xiong, W.}
\newblock \bibinfo{title}{Cavity-enabled enhancement of ultrafast
  intramolecular vibrational redistribution over pseudorotation}.
\newblock \emph{\bibinfo{journal}{Science}} \textbf{\bibinfo{volume}{378}},
  \bibinfo{pages}{790--794} (\bibinfo{year}{2022}).

\bibitem{LiJCP2020}
\bibinfo{author}{Li, T.~E.}, \bibinfo{author}{Nitzan, A.} \&
  \bibinfo{author}{Subotnik, J.~E.}
\newblock \bibinfo{title}{On the origin of ground-state vacuum-field catalysis:
  Equilibrium consideration}.
\newblock \emph{\bibinfo{journal}{J. Chem. Phys.}}
  \textbf{\bibinfo{volume}{152}}, \bibinfo{pages}{234107}
  (\bibinfo{year}{2020}).

\bibitem{MondalJCP2023}
\bibinfo{author}{Mondal, S.} \& \bibinfo{author}{Keshavamurthy, S.}
\newblock \bibinfo{title}{{Phase space perspective on a model for isomerization
  in an optical cavity}}.
\newblock \emph{\bibinfo{journal}{J. Chem. Phys.}}
  \textbf{\bibinfo{volume}{159}}, \bibinfo{pages}{074106}
  (\bibinfo{year}{2023}).

\bibitem{HoffmannJCP2020}
\bibinfo{author}{Hoffmann, N.~M.}, \bibinfo{author}{Lacombe, L.},
  \bibinfo{author}{Rubio, A.} \& \bibinfo{author}{Maitra, N.~T.}
\newblock \bibinfo{title}{{Effect of many modes on self-polarization and
  photochemical suppression in cavities}}.
\newblock \emph{\bibinfo{journal}{J. Chem. Phys.}}
  \textbf{\bibinfo{volume}{153}}, \bibinfo{pages}{104103}
  (\bibinfo{year}{2020}).

\bibitem{TichauerJCP2021}
\bibinfo{author}{Tichauer, R.~H.}, \bibinfo{author}{Feist, J.} \&
  \bibinfo{author}{Groenhof, G.}
\newblock \bibinfo{title}{{Multi-scale dynamics simulations of molecular
  polaritons: The effect of multiple cavity modes on polariton relaxation}}.
\newblock \emph{\bibinfo{journal}{J. Chem. Phys.}}
  \textbf{\bibinfo{volume}{154}}, \bibinfo{pages}{104112}
  (\bibinfo{year}{2021}).

\bibitem{ying_taylor_huo_2023}
\bibinfo{author}{Ying, W.}, \bibinfo{author}{Taylor, M.} \&
  \bibinfo{author}{Huo, P.}
\newblock \bibinfo{title}{Resonance theory of vibrational polariton chemistry
  at the normal incidence}.
\newblock \emph{\bibinfo{journal}{ChemRxiv}}  (\bibinfo{year}{2023}).

\bibitem{Sokolovskii2023}
\bibinfo{author}{Sokolovskii, I.}, \bibinfo{author}{Tichauer, R.~H.},
  \bibinfo{author}{Morozov, D.}, \bibinfo{author}{Feist, J.} \&
  \bibinfo{author}{Groenhof, G.}
\newblock \bibinfo{title}{Multi-scale molecular dynamics simulations of
  enhanced energy transfer in organic molecules under strong coupling}.
\newblock \emph{\bibinfo{journal}{Nat. Commun.}} \textbf{\bibinfo{volume}{14}},
  \bibinfo{pages}{6613} (\bibinfo{year}{2023}).

\bibitem{LinPRL2011}
\bibinfo{author}{Lin, N.}, \bibinfo{author}{Marianetti, C.},
  \bibinfo{author}{Millis, A.~J.} \& \bibinfo{author}{Reichman, D.~R.}
\newblock \bibinfo{title}{Dynamical mean-field theory for quantum chemistry}.
\newblock \emph{\bibinfo{journal}{Phys. Rev. Lett.}}
  \textbf{\bibinfo{volume}{106}}, \bibinfo{pages}{096402}
  (\bibinfo{year}{2011}).

\bibitem{cDMFT}
\bibinfo{author}{Park, H.}, \bibinfo{author}{Haule, K.} \&
  \bibinfo{author}{Kotliar, G.}
\newblock \bibinfo{title}{Cluster dynamical mean field theory of the mott
  transition}.
\newblock \emph{\bibinfo{journal}{Phys. Rev. Lett.}}
  \textbf{\bibinfo{volume}{101}}, \bibinfo{pages}{186403}
  (\bibinfo{year}{2008}).

\bibitem{KochPRB2008}
\bibinfo{author}{Koch, E.}, \bibinfo{author}{Sangiovanni, G.} \&
  \bibinfo{author}{Gunnarsson, O.}
\newblock \bibinfo{title}{Sum rules and bath parametrization for quantum
  cluster theories}.
\newblock \emph{\bibinfo{journal}{Phys. Rev. B}} \textbf{\bibinfo{volume}{78}},
  \bibinfo{pages}{115102} (\bibinfo{year}{2008}).

\end{thebibliography}
\bibliographystyle{naturemag}

\end{document}